\documentclass[12pt,a4paper,epsfig]{article}
\usepackage{amsfonts}
\usepackage{epsfig}
\usepackage{mathtools,slashed}
\usepackage{color}

\definecolor{azulon}{rgb}{0.,0.,0.8}
\definecolor{verde}{rgb}{0.1,0.5,0.1}
\definecolor{rojo}{rgb}{0.5,0.1,0.1}

\newcommand \sech{\mathop{\rm sech}\nolimits}

\title{{\bf Low energy dynamics of vibrating Kinks.}}

\author{J. Mateos Guilarte$^{(a)}$
\\ {\normalsize {\it $^{(a)}$ Instituto Universitario de F\'{\i}sica Fundamental y Matem\'aticas}}\\ {\normalsize {\it
Universidad de Salamanca, SPAIN}}
\date{}
}
\begin{document}
\maketitle
\section*{ABSTRACT}
Low energy dynamics of Kinks and Kink-AntiKink configurations in the Jackiw-Rebbi model is fully described. The strategy is based 
in the Collective Coordinates adiabatic approach. The necessary solution of Quantum Mechanical spectral problems, both for
scalar and spinorial wave functions, is unveiled as an intermediate step.


\section{Introduction}
Throughout the last 50 years a vast research activity in studying the dynamics of  topological defects experienced strong impetus. Given the relevance of topological defects in Fundamental/Mathematical Physics, Condensed Matter Physics, Cosmology, Biophysics and other branches of Science this task revealed itself as necessary. Except in Integrable Field Theories like sine-Gordon, Korteweg-de Vries, Kadomtsevt-Petviashvili equations, etcetera, no analitycal methods are appliable. Recently, however, numerical methos has been applied with sucess to analyze scattering of Kinks in several distinguished non integrable models that live in $(1+1)$-dimensions. We mention specifically the References \cite{Alonso-Izquierdo}-\cite{Queiroga}-\cite{Nieto} where numerical methods of integration have been applied to understand processes of scattering between Kinks and AntiKinks. Focusing in analyzing collisions between Kink shaped defects,  either simply travelling and/or wobling while travelling, succes in understanding their interactions emerged. Of course ,there is a vast Literature on this subject that can be found in the References just quoted. 
In a parallel attack to the understanding  of interactions between topological defects has been alternatively produced from
the adiabatic, low energy, side. Time dependence is restricted to the so called collective coordinates and the investigation is led to deal with dynamical systems with a finite number of degrees of freedom. Specifically, when the objects of research are Kinks, the central topic in this paper, the solution of the simplified system in Reference \cite{MaOlRoWe} shows astonishing qualitative agreement with the numerical results.

In one a priori completely different framework a big deal of research has been devoted to study how Fermions affect the dynamics
of systems in Field Theory and Condensed Matter Physics, see \cite{Jackiw}-\cite{Schrieffer}-\cite{Niemi1986}. In these papers 
a new phenomenon with far reaching consequences was discovered: the fractionization of the Fermi number in presence of topological defects, see also \cite{Fresneda} where the connection with index theorems, specifically the eta invariant, was explored. These findings aroused interest in studying physical settings where Fermions live in presence of topological defects as Kinks, see e.g. \cite{Vachaspati}-\cite{Bazeia}-\cite{Jaffe}.

In this work we shall concentrate in developing the collective coordinates adiabatic approximation when Fermions are present in the system.
A previous work in this line is the Reference \cite{Campos1} but our focus will be the paradigmatic Jackiw-Rebbi model, a simpler setting rich enough to obtain a great amount of information. More precisely, we shall continue in this paper the analysis developed in References \cite{Campos} and \cite {Bazeia1} to fully unveil the adiabatic dynamics of vibrating kinks and kink-antikink configurations.

To finish this Introduction we mention that a broader point of view is developed in two interesting papers on systems where fermions interact with kinks, \cite{Saad}-\cite{Weigel}. Quantum effects triggered by excited fermions on kinks are accounted for when a single fermionic bound state is taken into account.

\section{The Jackiw-Rebbi model in $\mathbb{R}^{1,1}$ Minkowski space-time}

Let us consider a QFT of fermions and bosons restricted to move on a line. The dynamics is governed by the action, (\ref{JRact}): 
\begin{eqnarray}
S_{JR}[\phi, \Psi]&=& \int_{\mathbb{R}^{1,1}} \, d^2x \, \left\{ \frac{1}{2}\partial_\mu \phi\partial^\mu \phi - \frac{\lambda^2}{2}(\phi^2-1)^2 + i \bar{\Psi}\gamma^\mu \partial_\mu\Psi -g \bar{\Psi} \phi \Psi \right\} \label{JRact} \\ \bar{\Psi}&=& \Psi^\dagger \gamma^0 \quad , \quad [\phi]=1 \, \, \, , \, \, \, [\Psi]=L^{-1/2} \, \, \, , \, \, [\lambda]=L^{-1}=[g] \nonumber
\end{eqnarray}
encompassing a quartic self-interaction of the bosons plus a Yukawa coupling between fermions and bosons. In the natural system of units where the Planck constant and the speed of light in vacuum are set to one, $\hbar=c=1$, the dimensions of the fields and couplings are shown below the action above.
The Jackiw-Rebbi Hamiltonian $H = H_{FB}+H_{B}$, (\ref{JR}), is in turn obtained via a Legendre transformation :
\begin{eqnarray}
 H_{FB} &=&  \int\, dx \, \Psi^\dagger(t,x)\left\{-i \alpha \frac{\partial}{\partial x} \right\} \Psi(t,x) +\int \, dx \, \Psi^\dagger(t,x)\left\{ g \phi(t,x) \beta\right\}\Psi(t,x)  \nonumber \\ H_B&=&\frac{1}{2} \int \, dx \, \left\{\Pi^2(x)+ \left(\frac{\partial\phi}{\partial x}\right)^2+\lambda^2\left(\phi^2(t,x)-1\right)^2 \right\} \label{JR}
\end{eqnarray}
The Dirac matrices $\alpha=\sigma_2$ and $\beta=\sigma_1$ are chosen like in \cite{Jackiw}. Here, $\sigma_1$ and $\sigma_2$ are Pauli matrices
and this choice corresponds to the Clifford algebra
\begin{eqnarray*}
&& \gamma^0=\sigma_1 \, \, , \, \, \gamma^1=i \sigma_3 \quad , \quad \gamma^5=\gamma^0\gamma^1=\sigma_2 \\
&&  [\gamma^\mu,\gamma^\nu]= 2 g^{\mu\nu}\quad , \quad g^{\mu\nu}={\rm diag}(1,-1) \, \, , \, \, \mu, \nu=0,1
\end{eqnarray*}

The Klein-Gordon and Dirac fields are maps from the Minkowski space respectively to the field of the reals and the fundamental irreducible representation of the ${\bf Spin}(1,1;\mathbb{R})$ group:
\[
\phi(t,x) = \,\, \mathbb{R}^{1,1} \, \, \longrightarrow \, \, \mathbb{R} \, \, , \, \, \, \, \Psi(t,x)=\left(\begin{array}{c} \psi_1(t,x) \\  \psi_2(t,x)  \end{array}\right): \, \, \mathbb{R}^{1,1} \, \, \longrightarrow \, \, {\it irr}_P {\bf Spin}(1,1; \mathbb{R}) \nonumber \, ,
\]
i.e., the transformations generated by $[\gamma^0,\gamma^1]$, and characterized by the Lorentz boost parameter $\chi$, acts on one spinor in the form
\[
 S_L[\chi]=e^{\frac{\chi}{4}[\gamma^0,\gamma^1]}=\left(\begin{array}{cc} \cosh \frac{\chi}{2} & i \sinh \frac{\chi}{2} \\ -i \sinh \frac{\chi}{2} & \cosh \frac{\chi}{2}\end{array}\right) \,  , \, \, \cosh \chi=\frac{1}{\sqrt{1-v^2}} \, \, , \, \,   \Psi_L(t,x)=S_L[\chi]\Psi(t,x) \nonumber
\]

The Euler-Lagrange (classical) field equations read:
\begin{eqnarray}
&& \Box \phi+ 2 \lambda^2 \phi(t,x)(\phi^2(t,x)-1)+g \bar{\Psi}(t,x)\Psi(t,x)=0 \label{hfieldeq}\\
&& i \gamma^\mu \frac{\partial \Psi}{\partial x^\mu}- g \phi(t,x)\Psi(t,x)=0 \label{ffieldeq} \, \, .
\end{eqnarray}
This system of coupled non-linear PDE's is very difficult to solve. The situation is better -and pertinent to the posterior canonical quantization of the system- if some static solution of the system of the form $(\phi_S(x), \Psi_S=0)$ is discovered:
\[
-\frac{d^2 \phi_S}{d x^2}+2 \lambda^2 \phi_S(x)(\phi_S^2(x)-1)=0
\]
In that case, one may search for solutions close to these static solutions
which linearize the (\ref{hfieldeq}-\ref{ffieldeq}) PDE system: 
 \begin{eqnarray}
 &&  \phi(t,x)=\phi_S(x)+\eta(t,x) \, \, \Rightarrow \, \, \Box\eta+2\lambda^2(3\phi_S^2(x)-1)\eta(t,x)={\cal O}(\eta^2) \label{hfieldeql}\\
 && i \gamma^\mu \frac{\partial\Psi}{\partial x^\mu}-g \phi_S(x)\Psi(t,x)={\cal O}(\eta\Psi)\label{ffieldeql}
 \end{eqnarray}
Since the terms in (\ref{hfieldeql}-\ref{ffieldeql}) with no derivatives of the fields are time inedependent it is convenient to solve the linear system via Fourier transform in time:
\begin{equation}
\eta(t,x)=\int_{-\infty}^\infty \frac{d \omega_B}{2\pi} \, e^{i \omega_B t}\eta(x;\omega_B) \quad , \quad \Psi(t,x)=\int_{-\infty}^\infty \, \frac{d\omega_F}{2\pi} e^{i \omega_F t} \Psi(x;\omega_F) \nonumber
\end{equation}
The linear PDE system (\ref{hfieldeql}-\ref{ffieldeql}) becomes equivalent to the spectral problems (\ref{scham}-\ref{dham}):
\begin{eqnarray}
&& h_{\rm Sch}\eta(x;\omega_B)=\Big[ -\frac{d^2}{d x^2}+2 \lambda^2(3 \phi_S^2(x)-1)\Big]\eta(x;\omega_B)=\omega_B^2 \eta(x;\omega_B) 
\label{scham} \\
&& h_{\rm D}\Psi(x;\omega_F)=\Big[ -i \alpha \frac{d}{d x}+g \beta \phi_S(x)\Big]\Psi(x;\omega_F)=\omega_F \Psi(x;\omega_F) \label{dham}
\end{eqnarray}
where $h_{\rm Sch}$ and $h_{\rm D}$ are respectively the quantum mechanical Schr$\ddot{\rm o}$dinger and Dirac operators in the background created by the $\phi_s$ classical solution.

\subsection{Bose/Fermi quanta in homogeneous field backgrounds}

The standard canonical quantization procedure to build the space of stationary states of $H$ and evaluate the quantum transitions within the Fock space states is based on finding the eigenwave functions of the Schr$\ddot{\rm o}$dinger operator and the eigenspinors of the Dirac operator to be taken as the one-particle states. The simplest solutions of the classical field equations are the two homogeneous, independent of $t$ and $x$, minima of the scalar potential energy whereas $H_{BF}$ is minimized by $\Psi_V=0$:
\begin{equation}
\phi_S(t,x)^\pm=\phi_V^\pm =\pm 1 \quad ,  \quad \Psi_S(x)=\Psi_V=0 \nonumber
\end{equation} 
Choice of one of these two configurations, e.g. $(\phi_S(t,x)^+=+1, \Psi_S(t,x)=0)$, as the ground state,  spontaneously breaks the  $\phi \to -\phi$ symmetry and the linear Klein-Gordon and Dirac equations become:
\begin{eqnarray}
 \frac{\partial^2 \phi}{\partial t^2}&=&\frac{\partial^2 \phi}{\partial x^2}- 4 \lambda^2 \phi(t,x) \label{eqKG} \\
 \left(\begin{array}{cc} i \frac{\partial}{\partial t} & 0 \\ 0 & i \frac{\partial}{\partial t}\end{array}\right) \cdot\left(\begin{array}{c} \psi_1(t,x) \\ \psi_2(t,x)\end{array}\right)&=& \left(\begin{array}{cc} 0 &-\frac{\partial}{\partial x}+g \\ \frac{\partial}{\partial x}+g & 0 \end{array}\right)\cdot\left(\begin{array}{c} \psi_1(t,x) \\ \psi_2(t,x)\end{array}\right) \label{eqDir}.
\end{eqnarray} 
Because there are no time and space dependent terms in the PDE operators in the linear equations (\ref{eqKG}) and (\ref{eqDir}) it is convenient to search for the general solution relying on  Fourier transform integrals, see (\ref{scawavpac}-\ref{spinwavpac})-\ref{spinwavpacp}){\footnote{We denote back $\eta(t,x)$ as $\phi(t,x)$ to follow the conventional notation.}}
\begin{eqnarray}
\phi(t,x)&=&  \int_{\mathbb{H}^+} \frac{d k}{\sqrt{4\pi \omega_B(k)}} \left[a(k)e^{-i\, \omega_B(k)\, t+i k x} + a^*(k) e^{i \, \omega_B(k)t-i k x}\right] \label{scawavpac}\\
\Psi(t,x)&=&\sqrt{g}\int_{\mathbb{H}^+} \frac{d k}{\sqrt{4\pi \omega_F(k)}} \left[b(k) u(k)e^{-i \omega_F(k) t+i k x} + c^*(k) v(k) e^{i \omega_F(k)t-i k x}\right] \label{spinwavpac}\\ \Psi^\dagger(t,x)&=&\sqrt{g}\int_{\mathbb{H}^+} \frac{d k}{\sqrt{4\pi \omega_F(k)}} \left[b^*(k) u^\dagger(k)e^{i \omega_F(k) t-i k x} + c(k) v^\dagger(k) e^{-i \omega_F(k)t+i k x}\right] \label{spinwavpacp}
\end{eqnarray}
where the integration is performed over the upper branches $\mathbb{H}_B^+$ and $\mathbb{H}_F^+$ of the hyperbolas: $\omega_B^2=k^2+4\lambda^2, \omega_B=+\sqrt{k^2+4 \lambda^2}$, and $\omega_F^2=k^2+g^2, \omega_F=+\sqrt{g^2+k^2}$.

In order to be the spinor expansion (\ref{spinwavpac}) the general solution of (\ref{eqDir}), $u(k)$ and $v(k)$ must be respectively the eigenspinors of the $2\times 2$-matrices with eigenvalues $\omega_F$ {\footnote{Note however that the spectral equation for $u$ is the same as the spectral equation for $v$ if $(\omega_F,k)$ is replaced by $(-\omega_F,-k)$. Notice also that it is possible to come back to $(\omega_F;k)$ provided that $g$ will be transmutted to $-g$, the essential property of antimatter.}}:
\begin{eqnarray*}
&& \left( \begin{array}{cc} 0 & g-i k \\ g+i k & 0 \end{array}\right)\cdot \left(\begin{array}{c} u_{1}(k) \\ u_{2}(k) \end{array}\right)= +\sqrt{k^2+g^2} \left(\begin{array}{c} u_{1}(k) \\ u_{2}(k) \end{array}\right) \\ && \left( \begin{array}{cc} 0 & -g-i k \\ - g+i k & 0\end{array}\right)\cdot \left(\begin{array}{c} v_{1}(k) \\ v_{2}(k) \end{array}\right)= + \sqrt{k^2+g^2} \left(\begin{array}{c} v_{1}(k) \\ v_{2}(k) \end{array}\right) \\ && u(k)= +\left(\frac{\sqrt{k^2+g^2}}{2 g}\right)^{1/2}\cdot\left( \begin{array}{c} 1 \\ \frac{g+i k}{\sqrt{k^2+g^2}}\end{array}\right) \quad , \quad  v(k)= \left(\frac{\sqrt{k^2+g^2}}{2 g}\right)^{1/2}\cdot\left( \begin{array}{c} \frac{- g-i k}{\sqrt{k^2+g^2}} \\ 1 \end{array}\right)
\end{eqnarray*}
which satisfy the standard orthonormality conditions:
\[
u^\dagger(k)u(k)=\frac{\omega_F}{g}=v^\dagger(k)v(k) \, \, , \, \, \, \bar{u}(k)u(k)=1=-\bar{v}(k)v(k)
\] 
together with $u^\dagger(k)v(-k)= 0$, which shows that $u(k)$ and $v(-k)$ are perpendicular. 

In the canonical quantization procedure the Fourier coefficients of the scalar field are promoted to creation and annihilation bosonic operators satisfying the commutationon rules:
\begin{equation}
[ \hat{a}(k_1),\hat{a}^\dagger(k_2)]=\delta(k_1-k_2) \quad , \quad [ \hat{a}(k_1),\hat{a}(k_2)]=0= [ \hat{a}^\dagger(k_1),\hat{a}^\dagger(k_2)]
\nonumber
\end{equation}
The ground state with no meson particles at all is annihilated by all the destruction bosonic operators
\begin{equation}
\hat{a}(k) \vert 0 \rangle_B =0 \, , \quad \forall k \nonumber
\end{equation}
Meson multiparticle states have the form (\ref{bfss}):
\begin{equation}
\prod_{j=1}^N\, [\hat{a}^\dagger(k_j)]^{n_j} \vert 0 \rangle_B= \vert n_1n_2 \cdots n_N \rangle \, \, , \quad n_j \in \mathbb{N} \, \, , \quad \sum_{j=1}^N n_j=N \label{bfss}
\end{equation}
and span the basis of the Bosonic Fock space, obtained via symmetric tensor product. 

Simili modo, the canonical quantization of the Dirac field courses via anticommutation relations (\ref{facr1}-\ref{facr2})
\begin{equation}
\{\hat{b}^\dagger(k_1), \hat{b}(k_2) \}=\delta(k_1-k_2) \quad , \quad \{\hat{c}^\dagger(k_1), \hat{c}(k_2) \}=\delta(k_1-k_2) \label{facr1}
\end{equation}
\begin{equation}
\{\hat{b}(k_1), \hat{b}(k_2) \}=0= \{\hat{c}(k_1), \hat{c}(k_2) \}  \label{facr2}
\end{equation}
Likewise the fermionic ground state  (\ref{fgs}) with neither electrons nor positrons is annihilated by all the
fermionic destruction operators
\begin{equation}
\hat{b}(k) \vert 0 \rangle_F=0=\hat{c}(k) \vert 0 \rangle_F \, \, , \quad \forall k \label{fgs}
\end{equation}
Electron/positron{\footnote{We shall refer to the Fermi quanta in the Jackiw-Rebbi model as electrons/positrons by analogy with QED. In the JR system there is no electric charge. The N$\ddot{\rm o}$ther's invariant associated to the $\mathbb{U}(1)$ symmetry is the Fermi number. }} multiparticle states (\ref{melec}-\ref{pelec}) form the basis of the Fermionic Fock space built from the one-particle states via antisymmetric tensor product:
\begin{equation}
\prod_{j=1}^N\, [\hat{b}^\dagger(k_j)]^{n^-_j} \vert 0 \rangle_F= \vert n^-_1n^-_2 \cdots n^-_N \rangle \, \, , \quad n^-_j = 0 \,{\rm or} \, 1 \, \, , \quad \sum_{j=1}^N n^-_j=N \label{melec}
\end{equation}
\begin{equation}
\prod_{j=1}^N\, [\hat{c}^\dagger(k_j)]^{n^+_j} \vert 0 \rangle_F= \vert n^+_1n^+_2 \cdots n^+_N \rangle \, \, , \quad n^+_j = 0 \, {\rm or} \, 1 \, \, , \quad \sum_{j=1}^N n^+_j=N \label{pelec}
\end{equation}
Quantization by anticommutators forces the antisymmetry of the multiparticle states and thus one state can only be either unnoccupied, $n_j^\pm=0$, or occupied only by one Fermion, $n_j^\pm=1$. Note that $n^+_j=1$ and $n^-_j=1$ is simultaneously possible describing one state 
with one electron and one positron, both with momentum $k_j$. 

\section{Bosonic and Fermionic Kink fluctuations}

Besides the the homogeneous static solutions this system admits also static space dependent solutions that, via Lorentz transformations, are travelling waves with Kink shape:
\begin{eqnarray*}
&& -\frac{d^2 \phi}{d x^2}\pm 2 \lambda^2 \phi(x)(\phi^2-1)=0 \, \, \, \Leftarrow \, \, \, \phi_K^\pm(\frac{x-v t}{\sqrt{1- v^2}}-a)= \pm \tanh [\lambda( \frac{x-vt}{\sqrt{1-v^2}}-a] \\ && -i \alpha \frac{d \Psi_K}{d x}+\beta(g \phi_K +i m \alpha )\Psi_K=0 \, \, \Leftarrow \, \, \, \Psi_K=0
\end{eqnarray*} 

Defining non dimensional space-time coordinates $\tau= \lambda t$, $y=\lambda x$ and considering small fluctuations 
\[
\phi(\tau,y)\simeq \phi_K(y)+ \eta(\tau,y) \quad , \quad \Psi(\tau,y)=0+ \psi(\tau,y)
\]
on the Kink classical background the expansion above is still solution of the field equations if the linear system of coupled PDE's
hold:
\begin{eqnarray*}
&& \left( \frac{\partial^2}{\partial \tau^2}-\frac{\partial^2}{\partial y^2}+4-\frac{6}{\cosh^2 y}\right)\phi(\tau,y)= {\cal O}(\phi^2) \\ && \left(i\frac{\partial}{\partial \tau}-i \sigma_2 \frac{\partial}{\partial y}+\nu \sigma_1 \phi_K(y)\right)\Psi(\tau,y)={\cal O}(\phi \Psi)
\end{eqnarray*}
Note that: (1) We choose $\Psi_K=0$ as Fermionic ground state and thus we neglected the Fermionic backreaction on the Kink at the classical level. (2) We introduce the important non dimensional parameter $\nu=\frac{g}{\lambda}$ that measures the strength of the Yukawa versus the scalar self-interaction couplings. (3) Again we abuse of notation in the linearized equations by writing $\phi(\tau,y)$ instead of $\eta(\tau,y)$ . Because there are no $\tau$-dependent terms in both operators it is natural the seaech of solutions via $\tau$-Fourier transform integrals:
\begin{eqnarray*}
&& \phi(\tau,y)= \int_{-\infty}^\infty \, d\tau \, e^{i \Omega_B \tau} \,  f_{\Omega_B}(y) \quad ,\quad \Omega_B=\frac{\omega_B}{\lambda}\\ && \Psi(\tau,y)=  \int_{-\infty}^\infty \, d\tau \, e^{i \Omega_F \tau} \, \psi_{\Omega_F}(y) \qquad, \qquad \Omega_F=\frac{\omega_F}{\lambda}
\end{eqnarray*}
such that the general solution of the linearized equations requires the solution of two quantum mechanical spectral problems,
one for a P$\ddot{\rm o}$sch-Teller/Schr$\ddot{\rm o}$dinger operator the other one for a Dirac operator in a Kink potential background.

 \subsection{Higgs bosons propagating over Kink topological defects}
 Starting wit the scalar/Bose case, the quantum mechanical spectral problem (\ref{pthams}) governing the scalar Kink fluctuations  reads:

 \begin{equation}
 h_{PT}f_B(y)=\left(-\frac{d^2}{ d y^2}+4 -\frac{6}{\cosh^2 y}\right) f_{\Omega_B}(y)=\Omega_B^2 f_{\Omega_B}(y)  \label{pthams} 
\end{equation}
 Fortunately the eigenvalues and eigenfunctions of this one-particle Hamiltonian are well known.  The discrete spectrum posseses two bound state eigenfunctions whose corresponding eigenvalues s are respectively $\Omega_B^2=0$ and $\Omega_B^2=3$., namely:
 
  1. \underline{Zero mode}
 \begin{equation}
 \Omega_0=0 \qquad , \qquad f_0(y)= \frac{1}{\cosh^2 y} \nonumber 
 \end{equation}
 This eigenfunction is due to the spontaneous breaking of the translational symmetry by the Kink.
 
2. \underline{Bound state: the shape fluctuation mode}
 \begin{equation}
 \Omega_{\sqrt{3}}^2=3 \qquad , \qquad f_{\sqrt{3}}(y)=\frac{\sinh y}{\cosh^2 y} \nonumber
 \end{equation}
 The next eigenfunction in the discrete spectrum responds to vibrations of the Kink, rather than translations,  with a frequency of $\sqrt{3}$ and it is referred to as shape mode because is accompanied by variations in the Kink shape{\footnote{This fluctuation mode is closely approximated by a Derrick mode obeying to a conformal dilatation, see Reference \cite{MaOlRoWe}. }}.
 
 Above these two bound fluctuation modes the eigenstates with energies over the threshold of the continuous spectrum $\Omega_B^2(0)=4$ arise.

3. \underline{Scattering states: the continuous spectrum}
 
\begin{equation}
\hspace{-0.3cm}\Omega^2_B(q)= q^2 +4 \, \,  , \quad f(y;q)= e^{i q y}(3 \tanh^2 y -3 i q \tanh y-(1+q^2)) = e^{i q y} P_2(\tanh y;q) \nonumber
\end{equation}
It is remarkable that the scattering involved is transparent, i.e, the reflection amplitude $r(q)$ is zero and the modulus of the transmission amplitude is one:
\[
t(q)=\frac{1-i q}{1+i q}\cdot \frac{2-i q}{2+ i q}  \, \, .
\]
From $t(q)$ the total phase shift and the spectral density are easily computed. The general solution (\ref{scafield}) is obtained as a linear superposition 
in terms of the eigenfunctions of the one-particle operator
\begin{eqnarray}
\phi(\tau,y)&=&\lim_{\varepsilon \to 0}\left(A_0 e^{-i \varepsilon \tau}+ A_0^* e^{i \varepsilon \tau}\right) f_0(y) +\left( A_{\sqrt{3}}e^{-i \sqrt{3}\tau}+A^*_{\sqrt{3}}e^{i\sqrt{3} \tau}\right)\cdot f_{\sqrt{3}}(y)\nonumber \\ &+& \int \, \frac{dq}{\sqrt{2 \sqrt{q^2+4}}}\left( A(q)e^{-i \sqrt{q^2+4} \tau}f(y;q)+ A^*(a) e^{i \sqrt{k^2+4} \tau}f^*(y;q) \right) \label{scafield}
\end{eqnarray} 

Canonical quantization courses as usual replacing the complex coefficients of the spectral expansion by quantum operators satifying commutative quantization relations:
\[
[\hat{A}_0, \hat{A}_0^\dagger ]=1 ,\quad , \quad [\hat{A}_{\sqrt{3}},\hat{A}^\dagger_{\sqrt{3}}]=1 \quad , \quad [\hat{A}(q_1),\hat{A}^\dagger(q_2)]=\delta(q_1-q_2)
\]
The differences with respect to  the Bosonic Fock space in the vacuum sector are three: (1) There is one state where a Boson is bounded to the Kink center travelling with  it at no cost of energy. (2) The shape mode is one state in the Fock space where one Boson is trapped by the Kink giving rise to one Kink excited state characterized by its vibration frequency. (3) There are many states where the Higgs quanta are scattered off the Kink but the outgoing particles wave escape from the Kink center as plane waves times Jacobi polynomias of order 2.

\subsection{Electrons/Positrons propagating over Kink topological defects}
Spinorial Kink fluctuations are determined from the spectral problem of the one-particle Kink-Dirac Hamiltonian (\ref{specfkf}) :
\begin{eqnarray}
&& h_{DK}\psi(y; \Omega_F)=\Omega_F \psi_(y;\Omega_F) \qquad , \qquad \psi(y;\Omega_F)=\frac{1}{\sqrt{g}}\left(\begin{array}{c}\psi_1(y;\Omega_F) \\ \psi_2(y; \Omega_f) \end{array}\right) \label{specfkf}\\
 && h_{DK}= \left(\begin{array}{cc} 0 & -\frac{d}{dy}+ \nu \tanh y  \\ \frac{d}{dy}+\nu \tanh y  & 0 \end{array}\right) \, \, ,\,\, \ \nu=\frac{g}{\lambda} \, \, \, , \, \, \, [\psi_1(y;\Omega_F)]=[\psi_2(y;\Omega_F)]=1 \nonumber \, .
\end{eqnarray}
Instead of solving directly the spectral problem  (\ref{specfkf}) we notice that the
 square of the Dirac-Kink operator is a diagonal matrix of contiguous in the hierarchy of P$\ddot{\rm o}$sch-Teller-Schr$\ddot{\rm o}$dinger operators.
 Moreover, defining the first-order differential operator $d_\nu=\frac{d}{d y}+\nu \tanh y$, the Darboux factorization method may be succesfully applied to find the spectrum. 
\begin{eqnarray*}
 h_{DK}^2 &=& \left(\begin{array}{cc} -\frac{d^2}{d y^2}+ \nu^2-\frac{\nu(\nu+1)}{\cosh^2 y} & 0 \\ 0 & -\frac{d^2}{d y^2}+ \nu^2- \frac{\nu(\nu-1)}{\cosh^2 y}\end{array}\right) \\ &=&\left(\begin{array}{cc} d_\nu^\dagger d_\nu & 0 \\ 0 & d_{\nu}d_{\nu}^\dagger \end{array}\right) = \left(\begin{array}{cc} d_\nu^\dagger d_\nu & 0 \\ 0 & d_{\nu-1}^\dagger d_{\nu-1}+2 \nu -1 \end{array}\right) 
\end{eqnarray*}

1. \underline{Fermionic zero modes}
Immediately one recognizes Fermionic zero modes and the non existent normalizable anti-Fermionic zero modes as living respectively
in the kermels of $d_\nu$ or $d_\nu^\dagger$. One finds:
\[
 h_{DK} \left(\begin{array}{c} \psi_1^{(0)}(y) \\ 0 \end{array}\right)=0 \, \, \Rightarrow \, \, \psi_1^{(0)}(y)=\frac{1}{\cosh^\nu y} \, \, \, , \, \, \,  h_{DK} \left(\begin{array}{c} 0 \\ \psi_2^{(0)}(y) \end{array}\right)=0 \, \, \Rightarrow \, \, \psi_2^{(0)}(y)=\cosh^\nu y
\]
Needless to say changing from Kink to anti-Kink the normalizable zero mode is the corresponding to antifermions

2. \underline{Fermionic Kink shape modes}
Focusing in the case when $g$ is a multiple of $\lambda$ and $\nu=N\in \mathbb{N}^*$ is a positive natural number,
there are $N-1$ proper bound states which correspond to vibrating Kink spinorial shape modes. The eigenvalues are well known
and show that these states carry imaginary momentum on the positive imaginary half-axis in the complex $q$-plane: $q=i\kappa_l${\footnote{In this case there is one half-bound state just at the threshold of the continuous spectrum with $l=N$. These \lq\lq half-states\rq\rq do not exist if $\nu \notin \mathbb{N}^*$. }}.
 
  \underline{Eigenvalues}:
\[
\Omega_F^{(l)}(\kappa_l;N)=\sqrt{(2N-l)l}=\sqrt{N^2-\kappa_l^2} \, \, , \, \, l=1,2, \cdots, N-1 \quad , \quad \kappa_l=N-l
\]
 
In terms of hypergeometric Gauss series truncated to polynomials the eigenspinors are also explicitly known.

\underline{Eigenspinors} $\psi^{(l)}(y)= \left(\begin{array}{c} \psi_1^{(l)}(y)\\  \psi_2^{(l)}(y)\end{array}\right)$
\[
\begin{array}{c}
\psi_1^{(l)}(y)=\frac{1}{\cosh^{N-l}y}\cdot {}_2F_1(2N+1-l, -l , N-l+1; \frac{1}{2}(1+\tanh y)) 
\\  \\ \psi_2^{(l)}(y)=\frac{1}{\cosh^{N-l} y}\cdot {}_2F_1(2N-l, -l+1 , N-l+1; \frac{1}{2}(1+\tanh y))  
\end{array}
\]
This identification has been possible because the bound state labelled by $l$ in the upper diagonal component and the bound state 
labelled by $j=l-1$ in the lower diagonal component of the square of the Dirac operator share identical eigenvalues. 

3. \underline{Fermions scattered off Kinks}

It remains to describe the spinorial fluctuations scattered off Kinks, i.e., not bounded to the Kink center. Of course, these fluctuations belong to the continuous spectrum of $h_{DK}$ (\ref{dkss}):
\begin{equation}
h_{DK} \psi(y;q)= \Omega_F(q) \psi(y,q) \qquad , \qquad \psi(y;q)=\left(\begin{array}{c} \psi_1(y;q)\\ \\ \psi_2(y;q)\end{array}\right)
\label{dkss}
\end{equation}
The eigenvalues are:

\underline{Eigenvalues}: $\Omega_F^2(q)=q^2+N^2 \, \, \Rightarrow \, \, \Omega_F(q)=+\sqrt{q^2+N^2}$

whereas the eigenspinors are proper Gauss hypergeometric series:

\underline{Eigenspinors}
\begin{eqnarray*}
 \psi_1(y;q)&=& A(q)({\rm sech} y)^{-i q} \cdot {}_2F_1[1-i q+N,- i q-N,1-i q ;\frac{e^y}{e^y+e^{-y}}] 
\\  \psi_2(y;q)&=& 
A(q)({\rm sech} y)^{-i q} \cdot {}_2F_1[-i q+N,- i q-N+1,1-i q ;\frac{e^y}{e^y+e^{-y}}] \label{scatwf2}
\end{eqnarray*}

Scattering scalar or spinor waves are characterized by their phase shifts and/or spectral densities. Considering the system defined on a finite interval of very large length $L$ with PBC the spectral densities of the scattering through the Kink wells suffered respectively by the upper and lower components are: 
\begin{eqnarray*}
&& \rho_F(q)=\rho_F^{(1)}(q)+\rho_F^{(2)}(q) \\
&& \rho_F^{(1)} =\frac{g L}{2\pi}+\sum_{j=1}^{N-1} \frac{j}{j^2+q^2}+\frac{N}{N^2+q^2} \quad , \quad  \rho_F^{(2)}=\frac{g L}{2\pi}+\sum_{j=1}^{N-1} \frac{j}{j^2+q^2}
\end{eqnarray*}
Note that besides the precise characterization of the continuous spectrum in terms of the wave number $q$ the bound states resurface as poles in the spectral density.

\subsection{Fermionic quanta}
To finish this Section we expand first
the classical spinor field in terms of the one-particle states:
\begin{eqnarray*}
\frac{1}{\sqrt{g}}\Psi^+(y) &=& B_0 \left( \begin{array}{c}{\rm sech}^N y \\ 0  \end{array} \right)+ \sum_{l=1}^{N-1} \, \Big[ B_l  \left(\begin{array}{c} \psi_1^{(l)}(y) \\ \psi_2^{(l)}(y) \end{array}\right)  e^{-i \Omega_F^{(l)} g t}+C_l^*\left(\begin{array}{c} \phi_1^{*(l)}(y) \\ \phi_2^{*(l)}(y) \end{array}\right)  e^{i \Omega_F^{(l)} g t} \Big] \\ &+& \int \frac{dq}{\sqrt{4\pi \Omega_F(q)}}\left[ B(q) \left(\begin{array}{c} \psi_1(y;q) \\ \psi_2(y;q) \end{array}\right)  e^{-i \Omega_F(q) g t}+  C^*(q) 
\left(\begin{array}{c} \phi^*_1(y;q) \\ \phi^*_2(y;q) \end{array}\right)  e^{i \Omega_F(q) g t} \right] 
\end{eqnarray*}
We stress that the eigenspinors $\psi$ of the Dirac-Kink operator and of its $g$ to $-g$ transformed $\phi$ have been taken as a complete system in the space of spinor fields.

 The next step is the promotion of the coefficients to Fermi operators demanding  anticommutation rules between them to establish the canonical quantization procedure:
\begin{eqnarray*}
&& \{\hat{B}_0, \hat{B}_0^\dagger\}=1 \qquad , \qquad \{ \hat{B}_{l_1}, \hat{B}^\dagger_{l_2}\}=\delta_{l_1 l_2} \qquad , \qquad \{ \hat{C}_{l_1}, \hat{C}^\dagger_{l_2}\}=\delta_{l_1 l_2} \\ && \{\hat{B}(q_1), \hat{B}^\dagger(q_2)\}= \delta(q_1-q_2)= \{\hat{C}(q_1), \hat{C}^\dagger(q_2)\} \, 
\, , \, \, \, 
\{ \hat{\Psi}^+(y_1), i \hat{\Psi}^{+\dagger}(y_2)\}=i \delta(y_1-y_2)
\end{eqnarray*}
The Fermi ground state and in general the Fermionic Fock space describing electron/positron multiparticle states with Fermi statistics built in it follows.

 As a practical computation we show the normal ordered Fermi number operator, all the annihilation operators placed at the right of the creation operators denoted by the $:\hat{F}:$ symbol.
\begin{eqnarray*}
:\hat{F}:&=& \int \, dx \, \frac{1}{2}\left[\hat{\Psi}^{+\dagger}(x), \hat{\Psi}^{+}(x)\right] \quad , \quad \sigma_1 \left(\begin{array}{c} \psi_1(y;q) \\ \psi_2(y;q) \end{array}\right)=\left(\begin{array}{c} \phi_1(y;q) \\ \phi_2(y;q) \end{array}\right)\\ 
: \hat{F} : &=& \frac{1}{2}[\hat{B}_0^\dagger, \hat{B}_0]+\sum_{l=1}^{N-1} (\hat{B}_l^\dagger \hat{B}_l-\hat{C}_l^\dagger\hat{C}_l)+\int dq \rho_F(q) (\hat{B}^\dagger(q)\hat{B}(q)-\hat{C}^\dagger(q)\hat{C}(q))\\ &=&\hat{N}_0-\frac{1}{2}+\sum_{l=1}^{N_1} (\hat{N}_l^--\hat{N}_l^+)+\int d q \rho_F(q) (\hat{N}^-(q)-\hat{N}^+(q))
\end{eqnarray*}
Due to the unpaired zero mode we see that the expectation value of this operator in any state of the Fermionic Fock space is fractionary. Note that because the $\sigma_1$ matrix maps the eigenspinors of $h_{DK}(g)$ into those of $h_{DK}(-g)$, not only the states in the discrete spectra are paired (except the zero mode) but also the spectral densities in the continuous spectra are identical.

 \section{Low energy dynamics of vibrating Kinks. Impact
 of Bosonic and Fermionic shape modes}

Study of the dynamics of topological defects in non-linear non-integrable field theories is an endeavour beyond the reach of analytical methods. Numerical analysis helped with more and more powerful computers has been sucessfully used to obtain information about this important subject because many types of topological defects exist in Nature. During the last twenty years of the XX Century an alternative route has been investigated by physicist and mathematicians. The idea is that at low energies the dynamics is essentially described by geodesic motion over the moduli space of these extended objects. More recently internal/shape vibrational modes of fluctuation are included in this effective finite dimensional dynamical systems. The degrees of freedom correspond to the collective coordinates of the defect, centers, phases and amplitudes of its vibrational modes. The adiabatic principle dictates that both the Kink moduli space coordinates and the shape mode amplitudes of Kink fluctuations support all the time dependence and describe the adiabtic evolution of the topological defect. The miracle is that sticking to this approximation, results obtained  by numerical methods, has been qualitatively confirmed in this simplified scenario.

In this Section our goal is to construct the effective adiabatic dynamics of the Kink collective coordinates encompassing both bosonic and fermionic shape fluctuation modes. We thus start with the kink solution but incorporate also the bosonic and fermionic shape
fluctuation modes. 

\subsection{Low energy dynamics of a single vibrating Kink}
In the simplest $g=2\lambda$ case, $\nu=2$, there is only a shape mode of each type and we focus on the configuration
\begin{eqnarray*}
\phi(\tau,y) & \simeq & \phi_K(y)+ A(\tau)\eta_{\sqrt{3}}(y)  = \tanh [y-a(\tau)]+ A(\tau) \frac{\sinh [y-a(\tau)]}{\cosh^2[y-a(\tau)]}\\
\frac{1}{\sqrt{g}}\Psi_{\sqrt{3}}(\tau,y,a,\Lambda)&=& \Lambda(\tau)\Phi(t,a(\tau))  \simeq  \frac{\Lambda(\tau)}{\cosh [y-a(\tau)]} \left(\begin{array}{c} \tanh [y-a(\tau)] \\ 1/\sqrt{3} \end{array}\right) 
\end{eqnarray*}
where the Kink configuration is supplemented by the bosonic and fermionc shape modes, both of frequency $\Omega=\sqrt{3}$.  The space 
parametrized by the collective coordinates is thus a four-dimensiona supermanifold.
There are two bosonic collective coordinates, the Kink center $a$ and the amplitude of the bosonic shape mode $A$, spanning the \lq\lq body\rq\rq. The \lq\lq soul\rq\rq of the submanifold is spanned in turn by one 
fermionic collective coordinate, the amplitude of the 
fermionic shape model: $\Lambda$.
 Now a very subtle point: in classical field theory bosonic fields present no problems. Classical Fermi fields are incompatible with the exclusion principle and thus, strictly speaking do not exist. A loophole to deal with this problem is to focus on the anticommutation rules and look at their classical limit, only satisfied by Grassman variables. It is then natural to consider classical Fermi fields as Grassmann fields but then there is no the possibility of having the Dirac sea as the ground state. One must cope with the existence of spinor waves propagating with negative energy.  Within this spirit, we shall consider that THE SHAPE MODE AMPLITUDE, $\Lambda= \Lambda_1 + i \Lambda_2$, is a complex Grassmann variable:
\begin{eqnarray*}
\Lambda^2&=&0=\Lambda^{*2} \quad , \quad \Lambda \Lambda^*+\Lambda^*\Lambda=0 \\ \Lambda_1^2&=& 0=\Lambda_2^2 \quad , \quad \Lambda_1\Lambda_2+\Lambda_2\Lambda_1=0 \quad , \quad \Lambda^*\Lambda=2 i \Lambda_1 \Lambda_2
\end{eqnarray*}

Under the adiabatic hypothesis where the temporal dependence of the system is encoded in the collective coordinates $a(\tau), A(\tau), \Lambda(\tau)$ the dynamics is reduced to a finite dimensional Lagrangian system. The kinetic energy and the potential energy receive contributions of both the bosonic and fermionic collective coordinates. Rememenbering the vibrating Kink configuration
\[
\phi_K(y,a,A)= \tanh (y-a)\left((1+\frac{A}{\cosh(y-a)}\right)
\]
and the two components of the Fermionic bound state of energy$\sqrt{3}$, the Fermionic shape mode of fluctuations over the vibrating Kink
\begin{eqnarray*}
 \psi^{(2)}_{1\sqrt{3}}(y,a,\Lambda)&=& \frac{\Lambda}{\cosh (y-a)}\cdot {}_2 F_1(4,-1,2,\frac{1}{2}(1+\tanh (y-a)) =\Lambda f(y,a )\\ \psi^{(2)}_{2\sqrt{3}}(y,a,\Lambda)&=& \frac{\Lambda}{\cosh (y-a)}\cdot {}_2 F_1(3,0,2,\frac{1}{2}(1+\tanh (y-a))=\Lambda g(y,a) 
\end{eqnarray*}
the effective Kinetic energy receives two contributions:
{\footnote{Note that the expressions for the Fermionic kinetic and potential energies are hermitian, because the anti-hermiticity of $\frac{\partial}{\partial\tau}$ and the anti-commutativity of the Grassman fields $\Psi^\dagger$ and $\Psi$. }}:
\begin{eqnarray*}
T_{\rm eff}^B &=& \frac{1}{2}\int_{-\infty}^\infty \, dy \, \left( \frac{\partial \phi_K}{\partial a} \dot{a} +\frac{\partial\phi_K}{\partial A} \dot{A}\right)^2= \frac{1}{2}\left[\left(\frac{4}{3}+\frac{\pi}{2}A+\frac{14}{15} A^2\right)\dot{a}\dot{a}+\frac{2}{3} \dot{A}\dot{A}\right] \\
T_{\rm eff}^F &=&-\Lambda^*\dot{\Lambda}\int_{-\infty}^{\infty} \, dy \, \left( f[y,a]^2 + g[y,a]^2\right)+ i \Lambda^* \Lambda 
\int_{-\infty}^{\infty} \, dy \, \left[ f[y,a]\frac{\partial f}{\partial a}+ g[y,a]\frac{\partial g}{\partial a}\right]\dot{a}=-
\frac{8}{3 }\Lambda^* \Lambda
\end{eqnarray*}
The effective potential energies due to the bosonic and fermionic collective variables are slightly more difficult to compute:
\begin{eqnarray*}
V^{B}_{\rm eff} &=& \frac{1}{2} \int_{-\infty}^\infty \, dy \, \left( \left( \frac{d \phi}{d y}\right)^2+\left(1-\phi^2\right)^2\right)=\left(\frac{4}{3}+A^2+\frac{\pi}{8}A^3+\frac{2}{35} A^4\right) \\
V_{\rm eff}^F &=& i \Lambda^* \Lambda\int_{-\infty}^\infty \, dy \, \left[ f(y,a)\left(-\frac{d}{d y}+ 2 \phi_K(y,a,A)\right) g(y,a)+ g(y,a)\left(\frac{d}{d y}+ 2 \phi_K(y,a,A)\right)f(y,a)\right] \\ &=& - i \Lambda^* \Lambda \left(4 +\frac{\pi}{2}A
\right) = i\Lambda^* \Lambda\cdot W[a,A]
\end{eqnarray*}

We notice now the main conceptual facts: (1) Fluctuations in the Kink  center of mass $a$ and the amplitude of vibrating shape modes, both scalar (bosonic) and spinorial (fermionic),  are entangled. Thus, if the Kink kinetic energy decreases the frequency of Kink oscillations increases. (2) The amplitude of the Fermionic fluctuations is coupled to the amplitude of the Bosonic ones via a Yukawa
interaction between one real and one complex degrees of freedom, remarckably independent on the Kink position $a$.
To make these statements more precise we look at  the motion equations derived from the effective Lagrangian
\[
L_{\rm eff}=T_{\rm eff}^B+T_{\rm eff}^F-V_{\rm eff}^B-V_{\rm eff}^{FB}  \, \, , \quad L_{\rm eff}^F=- \frac{4}{3} \Lambda^* \dot{\Lambda} -i W[a,A] \cdot \Lambda^*\Lambda
\]
\[
L_{\rm eff}^B=\frac{1}{2}\left[\left(\frac{4}{3}+\frac{\pi}{2}A+\frac{14}{15}A^2\right) \dot{a}\dot{a}+ \frac{2}{3}\dot{A}\dot{A}\right]+ \left(\frac{4}{3}+A^2+\frac{\pi}{8}A^3+\frac{2}{35}A^4\right) 
\]
which are, (\ref{cefcc}-\ref{aefcc}-\ref{grme}):
\begin{eqnarray}
&& \frac{d}{d \tau}\left[(\frac{4}{3}+\frac{\pi}{2}A+\frac{14}{15}A^2)\dot{a}\right]=i \frac{\partial W}{\partial a} \Lambda^* \Lambda
\label{cefcc} \\
&& \frac{2}{3}\ddot{A}=(\frac{\pi}{2}+\frac{28}{15}A)\dot{a}^2-(2 A+\frac{3\pi}{8}A^2+\frac{8}{25}A^3)-i\frac{\partial W}{\partial A} \Lambda^* \Lambda \label{aefcc}\\ && i \dot{\Lambda}^* =W[a,A]\cdot\Lambda^* \label{grme}
\end{eqnarray}
Consider first the situation where the Fermionic fluctuations are null: $\Lambda(\tau)=0, \forall \tau$. Then, the first of the EL equations gives rise to a constant of motion because $a$ is a ciclyc variable:
\[
(\frac{4}{3}+\frac{\pi}{2}A+\frac{14}{15}A^2)\dot{a}=C \, \equiv \, \, \dot{a}=\frac{C}{\frac{4}{3}+\frac{\pi}{2}A+\frac{14}{15}A^2} \, \, .
\]
Still in absence of fermionic fluctuations, and focusing in the regime of small shape mode amplitude the second motion equation becomes linear:
\begin{eqnarray*}
 && \ddot{A}\simeq D(C)-\omega^2(C)A +{\cal O}(A^2) \qquad , \qquad  D(C)= \frac{9\pi}{32} C^2  \\  && \omega^2(C)= 3 -\left( \frac{21}{20}-\frac{27 \pi^2}{256} C^2 \right)\, \, .
\end{eqnarray*}
Having chosen ,the $\Lambda=0=\Lambda^*$ solution of the motion equations for the Grassman degree of freedom the main impact of the shape mode in the Kink dynamics is clearly shown. The frequency of the Kink oscillations gets modified as a function of $C \propto \dot{a}$
i.e., the faster the vibrating Kink moves the lower the frequency of its oscillations becomes and viceversa. This transference from kinetic to 
potential energy in Kink dynamics is a semi-classical effect because the shape mode appears at order $\hbar$ in the $\hbar$ expansion 
of the Jackiw-Rebbi action. The dependence of the shape mode amplitude with time is easily recognized if Fermionic fluctuations do not enter the game
\[
A(\tau)=\frac{D(C)}{\omega^2(C)}+ {\cal A}\left(\cos (\omega(C) \tau ) + {\cal D}\right) \, \, .
\]
We stress that there exists a critical value of $C$. If $\vert C\vert > 2 \sqrt{\frac{5}{7}}$, $\omega(C)$ becomes imaginary and the Kink 
stop oscillating, the evolution is purely kinetic. When $\vert C\vert < 2 \sqrt{\frac{5}{7}}$ the translational and vibrational movements coexist. Any perturbation producing a variation of $C$ produces a variation of the shape mode frequency and viceversa, just qualitatively agreeing with the numerical predictions in the full field theoretical model.

The mutual influence between Bosonic and Fermionic fluctuations is understood if we consider the dynamics determined by the third equation. The equation (\ref{grme}) reads:
\begin{equation}
i \dot{\Lambda}^*(\tau)=-\left(4+ \frac{\pi}{2}A\right)\Lambda^* \label{locgeq}
\end{equation}

The formal solution to (\ref{locgeq}) is easy to find, i.e., see (\ref{grtds})
\begin{equation}
\Lambda^*(\tau)=\delta^*{\rm exp}[i \frac{1}{2}\int_0^\tau \, d\tau^\prime \, \left(4+\frac{\pi}{2}A(\tau^\prime \right)] \label{grtds}
\end{equation}
where $\delta^*=\delta_1-i \delta_2$, $\delta_1^2=\delta_2^2=0$, is a Grassmann integration constant. Thus, $i\Lambda^*(\tau)\Lambda(\tau)=i \delta^*\delta $ and therefore
\begin{eqnarray*}
&&  \ddot{A}\simeq D(C)-2\cdot  i\delta^*\delta-\left(\omega^2(C)-i\frac{\pi}{2} \delta^*\delta\right) A+{\cal O}(A^2) \, \, \Rightarrow \\ && A(\tau)\simeq\frac{D(C)-2\cdot i \delta^*\delta}{\omega^2(C)-i\frac{\pi}{2} \delta^*\delta}+{\cal A}\left( \cos ((\omega(C)-i \frac{\pi}{2} \delta^*\delta)\tau )+{\cal D}\right) \, \, .
 \end{eqnarray*}
The spinorial Kink fluctuations interact with the scalar Kink fluctuations modifying the midpoint of the scalar fluctuation amplitudes and the vibration frequencies.

In order to calibrate how the Kink dynamics depends on the oscillatory shape modes also for $\frac{g}{\lambda}=3$ let us consider the two vibrating modes.
The spinor fields describing these oscillations are:
\begin{eqnarray*}
\Psi^{(1)}_{\sqrt{5}}(\tau,y,a)&=&\Lambda(\tau)\Phi^{(1)}(y,a) =\frac{\Lambda(\tau)}{\cosh^2 (y-a(\tau))}\left(\begin{array}{c} 
{}_2F_1(6,-1,3,\frac{1}{2}(1+\tanh(y-a(\tau))\\ \, {}_2F_1(5,0,3, \frac{1}{2}(1+\tanh(y-a(\tau))\end{array}\right) \\ \Psi^{(2)}_{\sqrt{8}}(\tau,y,a)&=& \Lambda(\tau)\Phi^{(2)}_{\sqrt{8}}(y,a)=\frac{\Lambda(\tau)}{\cosh (y-a(\tau))}\left(\begin{array}{c} 
{}_2F_1(5,-2,2,\frac{1}{2}(1+\tanh(y-a(\tau))\\ \, {}_2F_1(4,-1,2, \frac{1}{2}(1+\tanh(y-a(\tau))\end{array}\right)
\end{eqnarray*}

Only the Fermionic kinetic and potential energies are modified because, even though $g/\lambda=3$, we stick to the standard $\phi^4$ Kink:
\begin{eqnarray*}
T^{\rm F(1)}_{\rm eff} &=&  - \Lambda^* \dot{\Lambda} \int_{-\infty}^\infty \, dy \, \left(\Phi^{T(1)}(y)\Phi^{(1)}(y) \right)=- \frac{8}{5} \Lambda^* \dot{\Lambda} \\ T^{\rm F(2)}_{\rm eff} &=&  - \Lambda^* \dot{\Lambda} \int_{-\infty}^\infty \, dy \, \left(\Phi^{T(2)}(y)\Phi^{(2)}(y) \right)=- \Lambda^* \dot{\Lambda } \\ V_{\rm eff}^{F(1)}&=& i \Lambda^* \Lambda \int_{-\infty}^\infty \, dy \, \Phi^{T(1)}(y)\left(-i \sigma_2\frac{d}{d y}+3 \sigma_1 \tanh y (1+\frac{A}{\cosh y})\right)\Phi^{(1)}(y)=- i \Lambda^* \Lambda \, \left(\frac{8}{3}+\frac{3\pi}{8} A \right) \\   V_{\rm eff}^{F(2)}&=& i \Lambda^*   \Lambda \int_{-\infty}^\infty \, dy \, \Phi^{T(2)}(y)\left(-i\sigma_2 \frac{d}{d y} +3 \sigma_1\tanh y (1+\frac{A}{\cosh y})\right)\Phi^{(2)}(y) =-i\Lambda^* \Lambda \left(\frac{8}{3}+\frac{21}{64} A\right)
\end{eqnarray*} 

In sum, the effective models are Lagrangian dynamical systems where the configuration spaces  are supermanifolds. The dynamical variables are both $c$-umbers, $a$ and $A$, and Grasmann magnitudes, $\Lambda^*$ and $\Lambda$. The fact that the effective super dynamical systems are not supersymetric is due to the fact that the parent field theory, the Jackiw-Rebbi model is not supersymmetric. {\it Although we started with no backreaction of the fermions on the Kink at the classical level this effect has been generated at one-loop level by taking into account the effect of Fermionic shape modes}

\subsection{ Effective dynamics of vibrating Kink-Anti-Kink configurations}

\subsubsection{Effective kinetic energy}
The Kink-Antikink configurations depend on two parameters when the two basic topological defects are vibrating :
\begin{eqnarray}
\phi_{KA}(y,a(\tau),A(\tau))&=&\tanh (a(\tau)+y)-\tanh(y-a(\tau)) -1 \label{KAvibad} \\ &+&\frac{ A(\tau)}{\tanh (a(\tau))}  (\tanh
   (a(\tau)+y)\cdot
   \text{sech}(a(\tau)+y)-\tanh (y-a(\tau))\cdot
   \text{sech}(a(\tau)-y)) \nonumber
\end{eqnarray}
where $a$ labels the relative position of the Kink with respect to the AntiKink and $A$ the syncronized amplitudes of vibration of Kink and Antikink. It is assumed that the Center of Mass is placed in the origin of the Reference system.

In the formula (\ref{KAvibad}) the adiabatic approximation is implemented: only the collective coordinates $a$ and $A$ depend on the $\tau$ time. The evoluion is so smooth that the spatial coordinate $y$ remains constant in time. Under this hypothesis a non Euclidean  metric arises in the $(a,A)$ plane:
\begin{eqnarray*}
g_{aa}(a,A)&=&\int_{-\infty}^\infty \, dy \, \left(\frac{\partial\phi_{KA}}{\partial a}\cdot \frac{\partial\phi_{KA}}{\partial a}\right) \, \, \, , \, \, \, g_{aA}(a,A)=\int_{-\infty}^\infty \, dy \, \left(\frac{\partial\phi_{KA}}{\partial a}\cdot \frac{\partial\phi_{KA}}{\partial A}\right) \\ g_{AA}(a,A)&=&\int_{-\infty}^\infty \, dy \, \left(\frac{\partial\phi_{KA}}{\partial A}\cdot \frac{\partial\phi_{KA}}{\partial a}\right) \quad , \quad \dot{a}=\frac{\partial a}{\partial \tau} \, \, , \, \, \dot{A}=\frac{\partial A}{\partial \tau}
\end{eqnarray*}
such that the Kinetic energy of the Kink-AntiKink adiabatic evolution becomes:
\[
T_{\rm eff}^{KA}=\frac{1}{2} \left(g_{aa}(a,A)\dot{a}^2+2 g_{aA}(a,A) \dot{a}\dot{A }+g_{A,A}(a,A)\dot{A}^2  \right)
\]
Mathematica computations offer the following results
\footnotesize{\begin{eqnarray*} &&g_{aa}(a,A)=\\ &=&\frac{480 a e^{10 a} \left(A^2
   \cosh (8 a)+4 \left(46
   A^2+7\right) \cosh (2 a)+4
   \left(23 A^2-4\right) \cosh
   (4 a)+ 4 \left(2
   A^2+1\right) \cosh (6
   a)+163 A^2-16\right)}{15 \left(e^{2
   a}-1\right)^7 \left(e^{2
   a}+1\right)^3}\\&-& \frac{16
   e^{10 a} \sinh (2 a)
   \left(40 \left(93
   A^2-13\right) \cosh (2
   a)+\left(668 A^2+80\right)
   \cosh (4 a)+40 \left(3
   A^2+1\right) \cosh (6
   a)\right)}{15 \left(e^{2
   a}-1\right)^7 \left(e^{2
   a}+1\right)^3}\\ &-& \frac{16 e^{10 a}\sinh (2a)\left(\left(7 A^2+10\right)
   \cosh (8 a)+2219
   A^2+410\right)+15 \pi 
   \left(e^{2 a}-1\right)^{10}
   A}{15 \left(e^{2
   a}-1\right)^7 \left(e^{2
   a}+1\right)^3}
\end{eqnarray*}}

\footnotesize{\begin{eqnarray*}
g_{aA}(a,A)&=& -A \tanh (a)-5 a A
   \text{csch}^6(a)-7 a A
   \text{csch}^4(a)-3 a A
   \text{csch}^2(a)+(\pi -a A)
   \text{sech}^2(a)\\ &+&\coth (a)
   \left(5 A
   \text{csch}^4(a)+\frac{11}{
   3} A
   \text{csch}^2(a)+A\right)
\end{eqnarray*}}

\footnotesize{\[
g_{AA}(a)=\frac{1}{24} \text{csch}^5(a)
   \text{sech}(a) (36 a-3
   \sinh (2 a)-12 \sinh (4
   a)+\sinh (6 a)+12 a \cosh
   (4 a))
\]}
Starting with the $aa$ component of the metric tensor we observw that the limits when $a$ tends to $\pm \infty$ are
\[
\lim_{a \to \pm \infty} g_{aa}[a,A]=\frac{8}{3}\pm \pi A +\frac{28}{15} A^2 \, \, \quad .
\]
These limits correspond to infinite separation between the Kink and AntiKink centers, the AntKInk at the right with respect to the Kink, the $+$ sign, or viceversa, the $-$ sign. Clearly, this component of the metric tensor is twice the metric of one excited single Kink. Near the origin, the $aa$ component metric tensor behaves as follows:
\[
g_{aa}[a,A] \simeq_{a \to 0} \pi  a^3 A+a^2 \left(\frac{248
   A^2}{63}-\frac{64}{15}\right)+\frac{16}{3}
\]

The limits of the other components of the metric tensor at $\pm \infty$ are
\[
\lim_{a \to \pm \infty} g_{aA}[a,A]=0 \quad , \quad \lim_{a  \to \pm \infty}g_{AA}`[a,A]=\frac{4}{3}
\]
confirming that when Kink AntiKink are very far appart behave as two isolated single Kinks and their interactions are negligible. Close to the origin when Kink and AntiKink are partially superposed these components of the metric tensor become:
\[
g_{a A}[a,A] \simeq_{a \to 0} -\pi a (\frac{152}{105} A + a)+\frac{80}{63} A a^3 \quad , \quad g_{A A}[a,A] \simeq_{a \to 0} \frac{56}{15}-\frac{152}{105} a^2
\]

\begin{figure}[ht]
\centerline{\includegraphics[height=3.5cm]{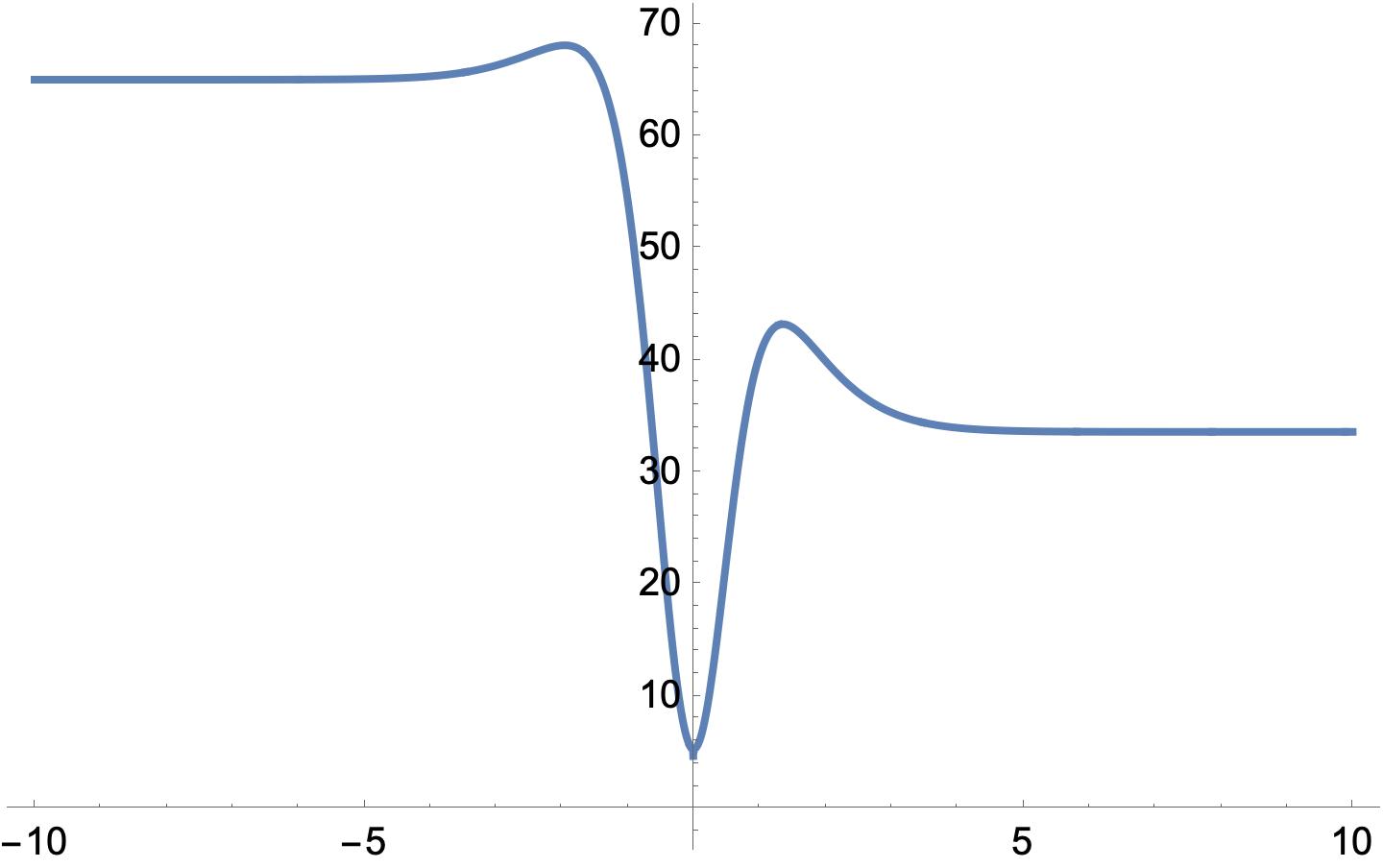} \hspace{0.2cm}\includegraphics[height=3.5cm]{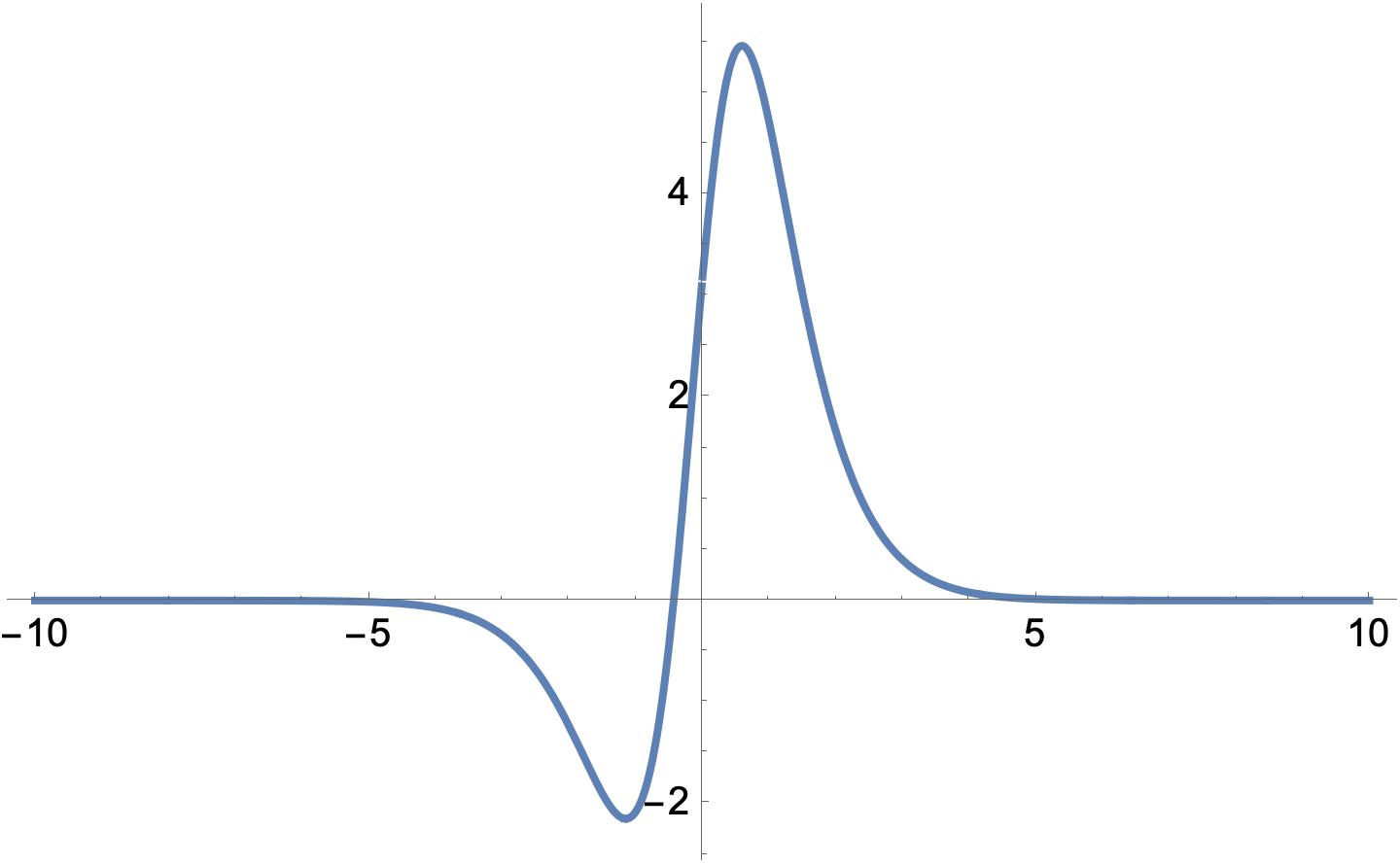}\hspace{0.15cm} \includegraphics[height=3.5cm]{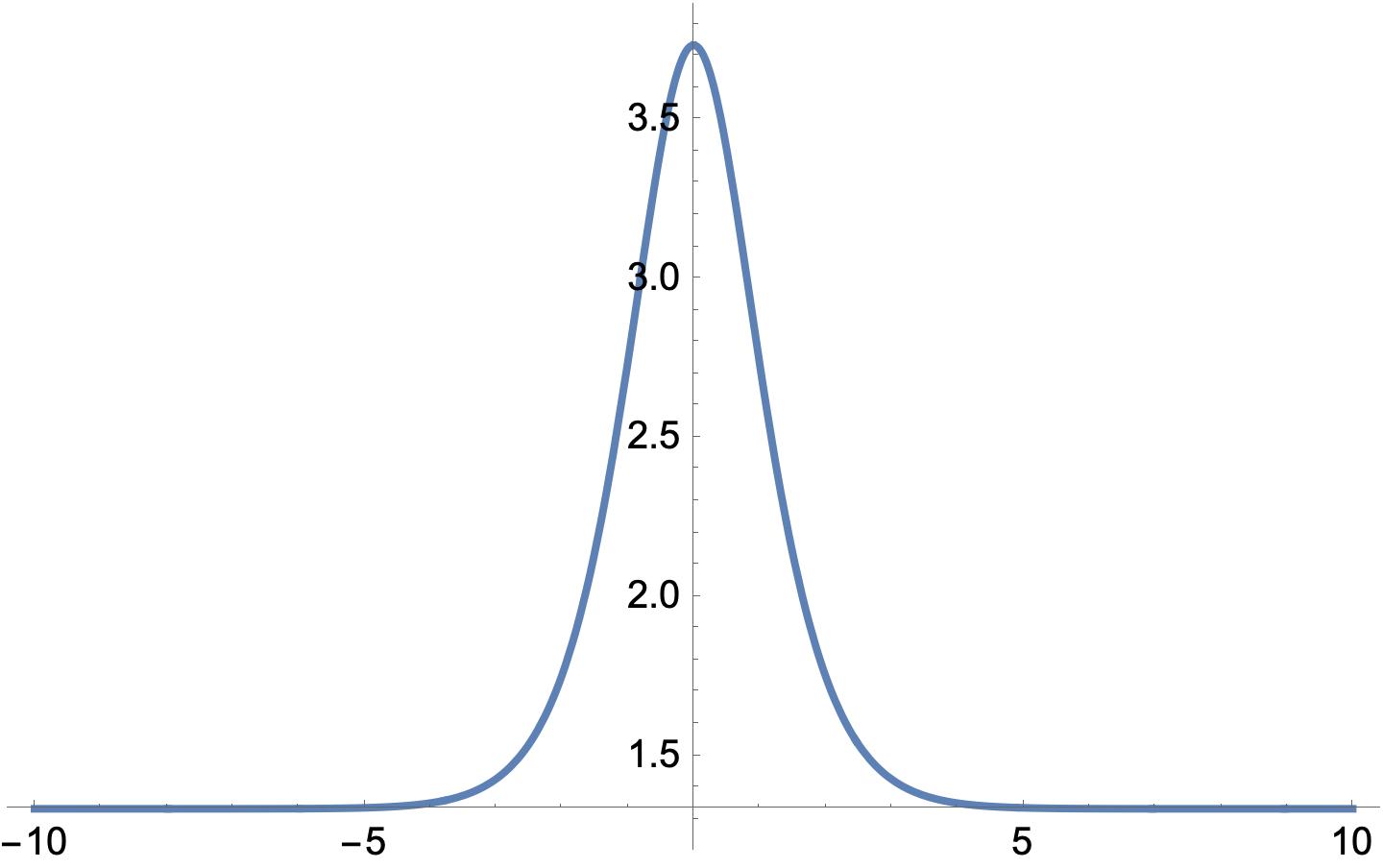}}
\caption{Snapshots of: (left) $g_{aa}[a,-5]$ (center) $g_{aA}[a,5]$ (right) Graphics of $g_{AA}[a]$ } 
\end{figure}

\subsubsection{Effective potential energy}

The last step is the computation of the effective potential energy between vibrating Kink and Antikink topological defects in the
adiabatic regime of collective coordinates, (\ref{leeffpot}):
\begin{equation}
V_{\rm eff}^{KA}[a,A]= \frac{1}{2}\int_{-\infty}^\infty \, dy \, \left(\frac{\partial \phi_{KA}}{\partial y}\cdot\frac{\partial \phi_{KA}}{\partial y}+\left(1-\phi^2_{KA}(y,a,A)\right)^2\right) \label{leeffpot}
\end{equation} 
The calculation needs a huge computational effort and here is the result achieved in a Mathematica enviroment:
\footnotesize{\begin{eqnarray*}
V_{\rm eff}^{KA}[a,A]&=&\frac{1}{210 (e^{2 a}
-1)^{11}(e^{2 a}+1)^3} \times \left\{ \frac{105}{2} \pi  \left(e^{2
   a}-1\right)^{11} A
   \left(\left(-141 e^{2 a}+45
   e^{4 a}+e^{6 a}-1\right)
   A^2+96 \left(e^{2
   a}-1\right)\right)\right. \\ &&  \hspace{-2.2cm} +  16
   \left(e^{4 a}-1\right)
   3 e^{12 a} A^4 (10696
   \cosh (2 a)-15105 \cosh (4
   a)+2716 \cosh (6 a)-986 \cosh (8 a)+28 \cosh (10 a)+\cosh (12 a)-17510)-  \\ && \hspace{-1.8cm} -840
   e^{12 a} A^2 \sinh^4(a)
   (-288 \sinh (2 a)+40 \sinh
   (4 a)-96 \sinh (6 a)+4
   \sinh (8 a) + \\ && \hspace{-1.8cm}  +  656 \cosh (2
   a)-124 \cosh (4 a)+112
   \cosh (6 a)-5 \cosh (8
   a)-159)+(35 \left(-8 e^{4
   a}+e^{8 a}-17\right)
   \left(e^{2
   a}-1\right)^8 +\\ &+& \hspace{-1.8cm}
   1680 a \left(8 e^{11 a} \left(3
   e^{3 a} A^4 (130 \cosh (2
   a)-32 \cosh (4 a)+29 \cosh
   (6 a)-4 (\cosh (8 a)+7)+\cosh (10 a) )+ \right. \right.\\ && \hspace{-1.8cm} +12 e^{3
   a} A^2 \sinh ^4(a) (8 \sinh
   (2 a)-80 \sinh (4 a)+8
   \sinh (6 a)-8 \sinh (8
   a)- \\ && \hspace{-1.5cm}- 80 \cosh (2 a)+124 \cosh
   (4 a)- 16 \cosh (6 a)+9
   \cosh (8 a)+123)+\\ && \hspace{-1.5cm} +11 (42
   \sinh (a)-5 (6 \sinh (3
   a)-3 \sinh (5 a)+\sinh (7
   a))+ \sinh (9 a))+\\ && \left. \left. \left. \hspace{-1.5cm} + 14 \cosh
   (a)-22 \cosh (3 a)+5 \cosh
   (5 a)+7 \cosh (7 a)-5 \cosh
   (9 a)\right)+8\right)\right\}
\end{eqnarray*}
Close to the origin the effective Kink-AntiKink potential behaves as
\begin{eqnarray*}
V_{\rm eff}^{KA}[a,A] && \simeq_{ a \to 0} \frac{1}{420} a^3 \left(-3255
   \pi  A^3+11520 A^2+1680 \pi
    A-7168\right)+\\ && +\frac{a^2
   \left(-94912 A^4+45045 \pi 
   A^3+322036 A^2-180180 \pi 
   A+192192\right)}{15015}+\\ && + \frac{2}{35} a \left(105 \pi 
   A^3-608 A^2+105 \pi 
   A\right)+\frac{5152
   A^4-3465 \pi  A^3+15444
   A^2}{1155}
\end{eqnarray*}
Special values of the effective potential at distinguished points are:
\[
\lim_{a \to 0} V_{\rm eff}^{KA}[a,A]=\frac{736 A^4}{165}-3 \pi 
   A^3+\frac{468 A^2}{35} \quad , \quad \lim_{a \to \infty}V_{\rm eff}^{KA}[a,A]=\frac{8}{3}+\frac{\pi}{4}A^3+\frac{4}{35}A^4 \quad , \quad \lim_{a \to -\infty}V_{\rm eff}^{KA}[a,A]=\infty
\]
The qualitative properties of this mechanical potential are encoded in Figures 2 and 3.

\begin{figure}[ht]
\centerline{\includegraphics[height=3cm]{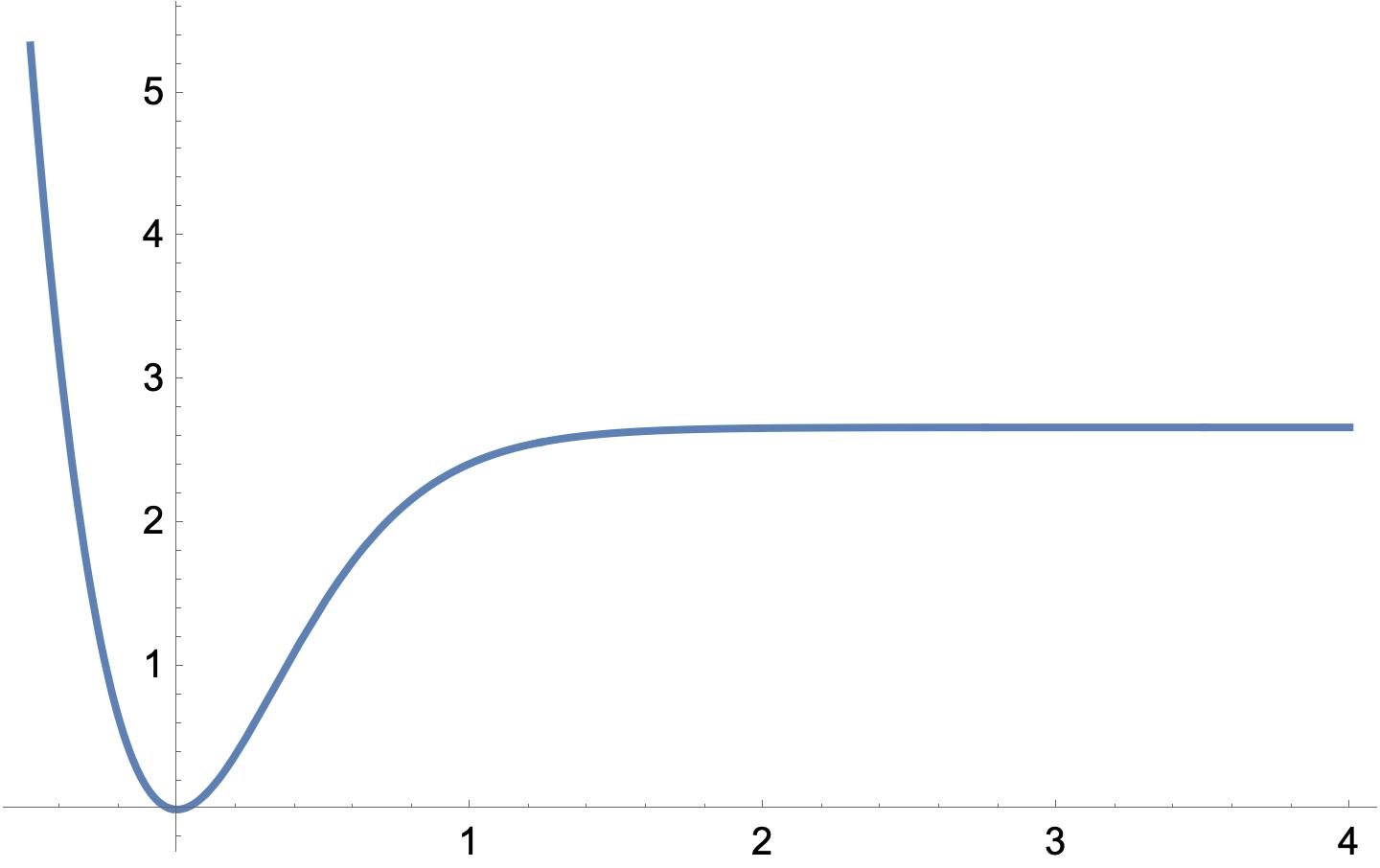} \hspace{1cm} \includegraphics[height=3cm]{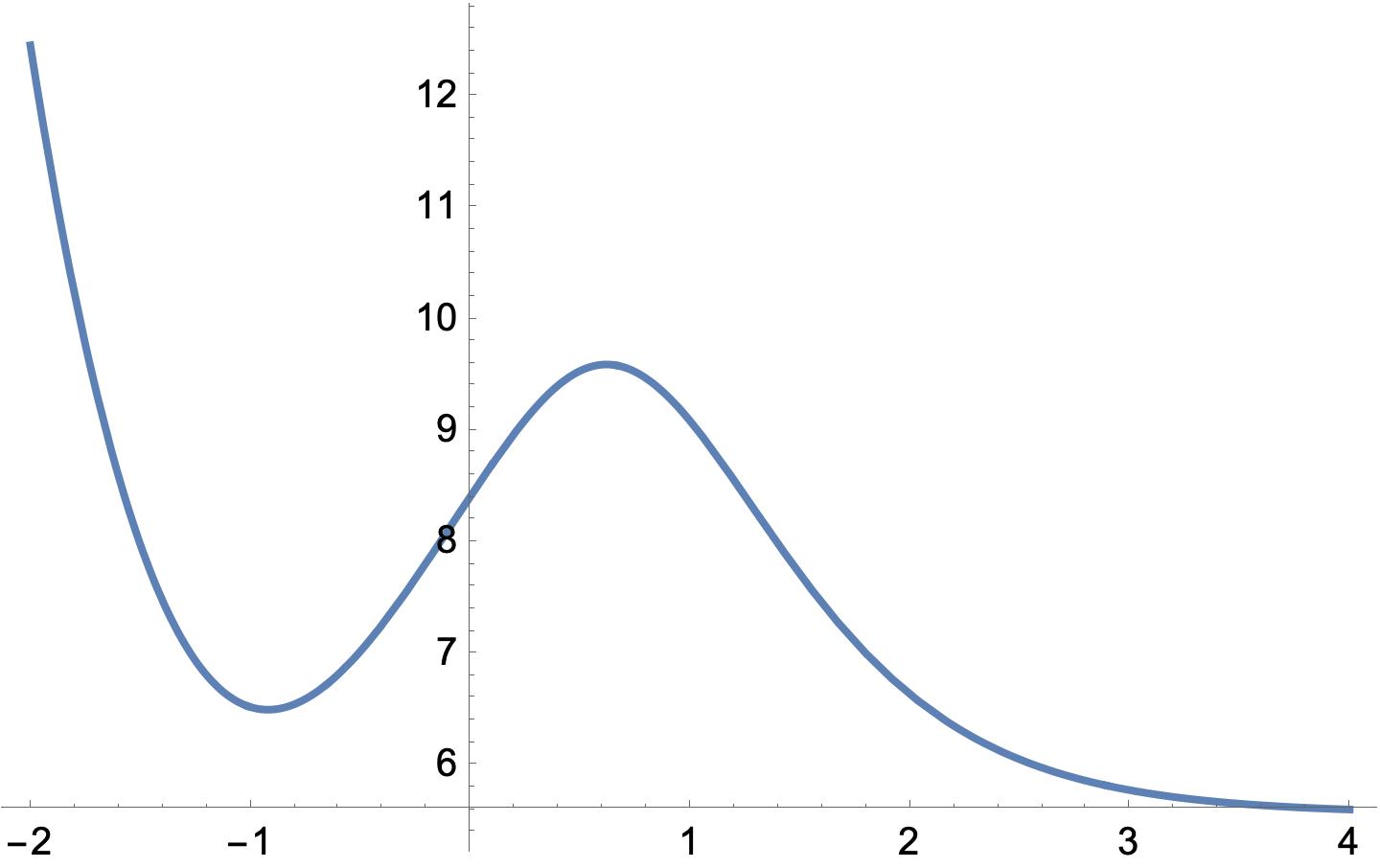} \hspace{1cm} \includegraphics[height=3cm]{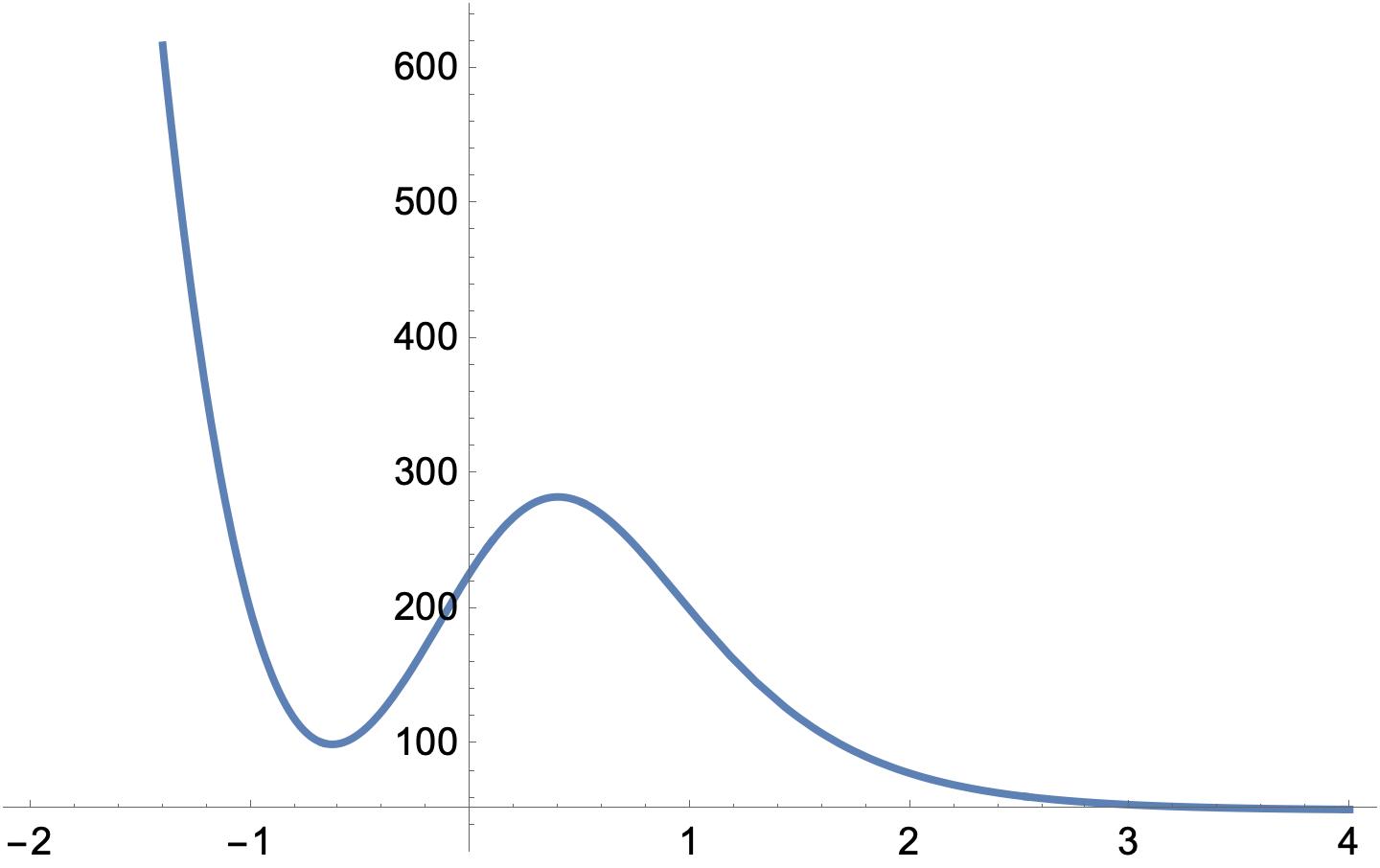} }
\caption{Tomographic snapshots of the effective potential when the Kink and the Anti-Kink are close to each other for the following  amplitudes of the shape mode: $A=0$ (left) $A=1$ (center) $A=3$ (right) } 
\end{figure}

\begin{figure}[ht]
\centerline{\includegraphics[height=6cm]{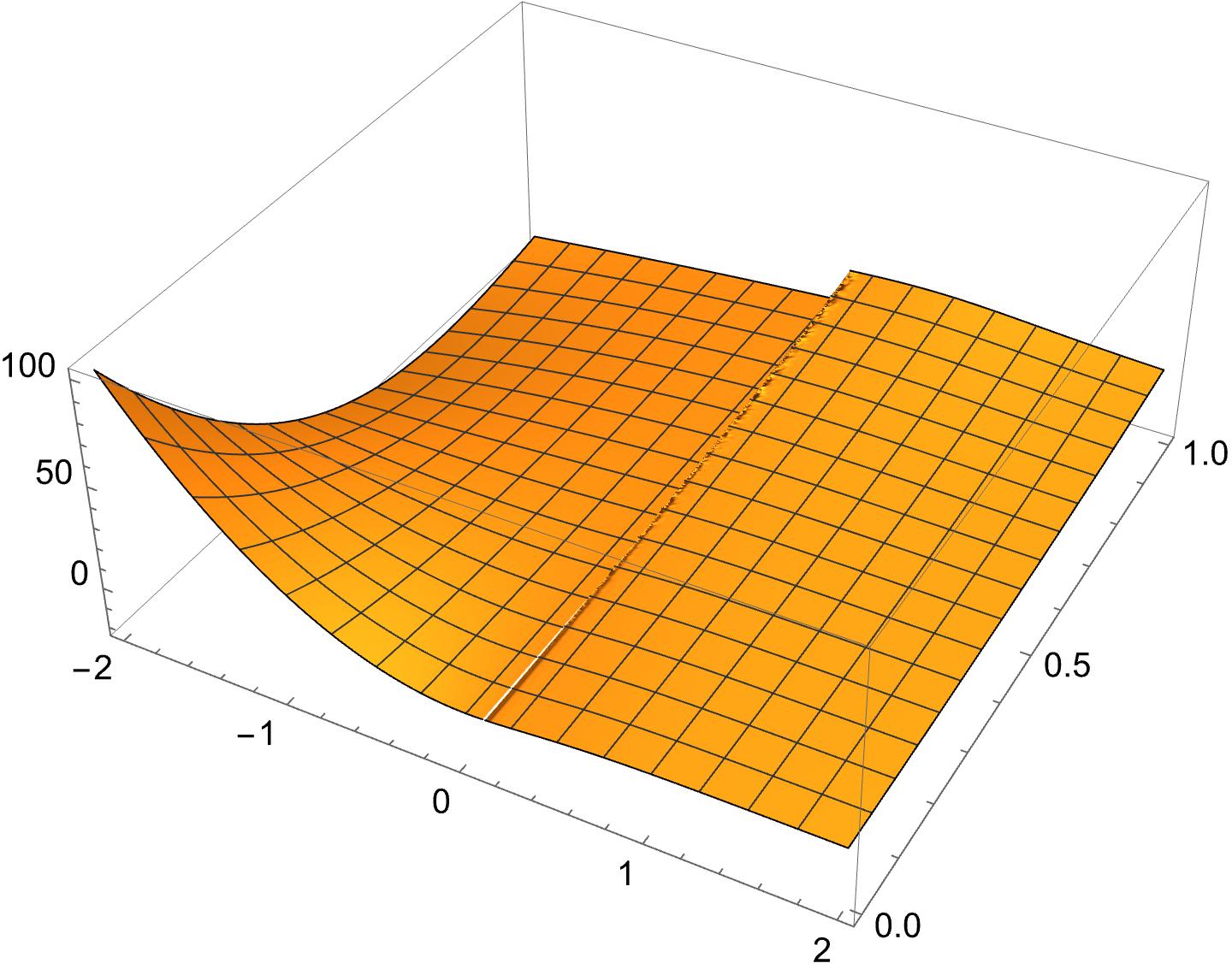} \hspace{2cm} \includegraphics[height=6cm]{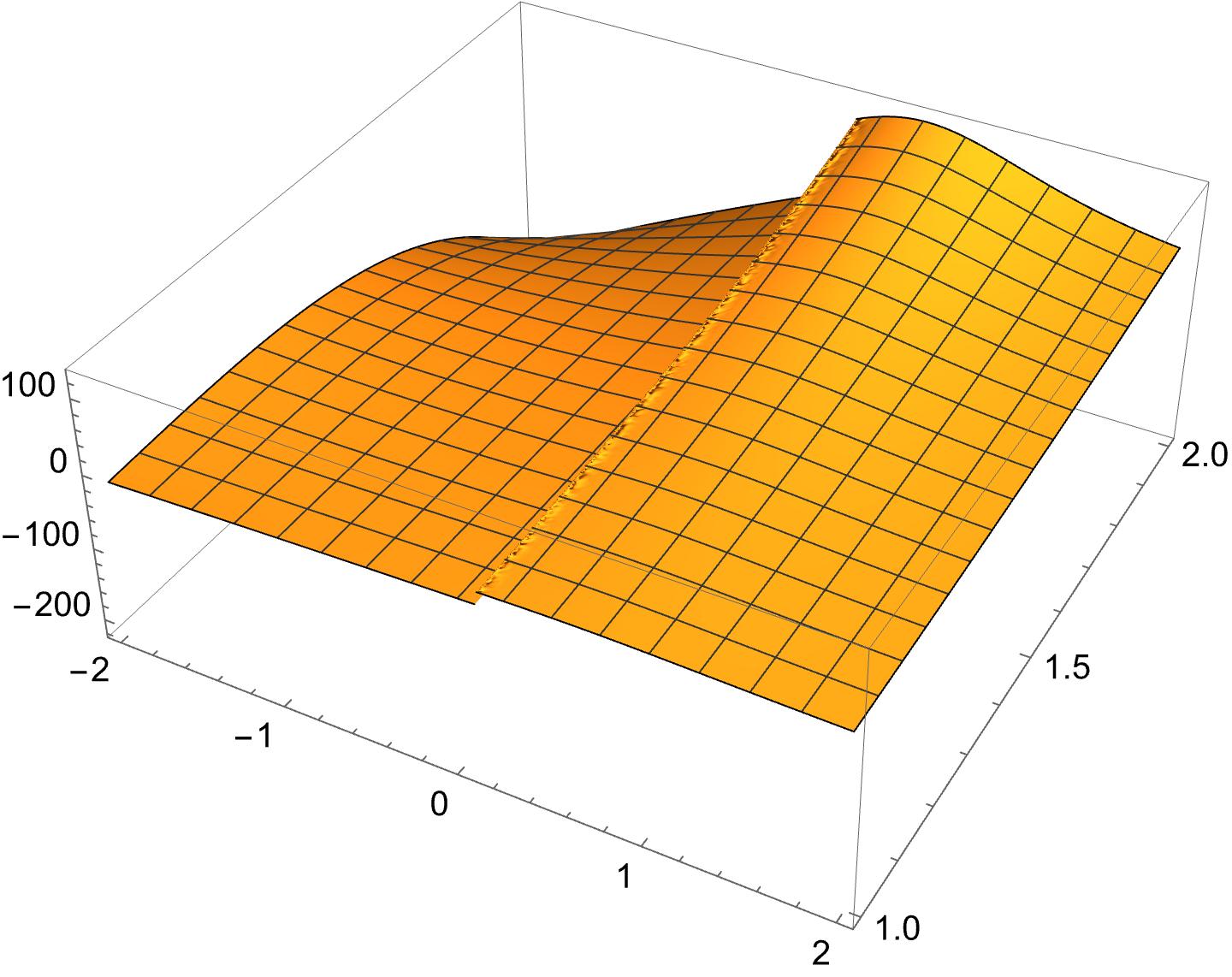}}
\caption{3D graphics of the effective potentialin the $(a,A)$-plane as function of the $(a,A)$-collective coordinates. Interval: $a\in (-2,2)$, $A\in (0,1)$ (left) $a\in (-2,2)$, $A\in(1,2)$ (right) \, .}
\end{figure}

If both Kink and AntiKink are non excited, i.e., when $A=0$, the low energy dinamics is captured by a Lagrangian mechanical system with a single degree of freedom, the relative position $a$:
\[
L= \frac{1}{2}g[a,0] \dot{a}\dot{a}- V_{\rm eff}^{KA}[a,0] 
\]
Therefore, the system is Liouville integrable and, because the energy
\[
E=\frac{1}{2} g[a,0]\dot{a}\dot{a} +V_{\rm eff}^{KA}[a,0]
\]
is a constant of motion, it is possible to reduce the integration of the system to the quadrature (\ref{quadt}):
\begin{equation}
t-t_0= \int \, da \, \sqrt{\frac{g[a,0]}{2(E- V_{\rm eff}^{KA}[a,0])}} \label{quadt}
\end{equation}
It is not possible neither writing the integral in terms of analytical functions nor, even if it not were the case, to invert the outcome 
and to know the explicit dependence of $a$ on time. Nevertheless, Figure 2.(left) as well as the explicit knowledge of 
$V_{\rm eff}^{KA}[a,0]$ makes possible a qualitative analysis of the motion.
 $E=\frac{8}{3}$ is the threshold for unbounded motion and bounded motion occurs if $0< E <\frac{8}{3}$. Thus,
 for energies greater than $\frac{8}{3}$, starting with initial conditions $(a(t-t_0<< 0) >> 0, \dot{a}(t-t_0<<0)< 0)$ $a$ decreases when times runs forward
until it becomes slightly negative meaning that Kink and AntiKink centers cross each other while exchanging their relative position.
For suficciently high energy, the relative position $a$ becomes negative enough to reach sooner or latter the infinite potential barrier. Then a bounce is produced and a second exchange between Kink and AntiKink takes place moving appart again the KAK pair up to very long distances. For energie less than $\frac{8}{3}$ but greater than $0$ the Kink-Antikink motion is bounded. These oscillatory motions were christened as \lq\lq bions\rq\rq  by its dicoverers in References \cite{Sugiyama}- \cite{Campbell}. The plots of the effective potentials for $A=1$ and $A=3$, Figure 2 (center) and (right) show similar pattern but less room for bions is left with increasing shape mode amplitudes.

A brief digression on the quantum description of the previously described classical dynamics  is convenient. Because the momentum conjugate to $a$ 
is $p_a=g[a,0]\dot{a}$ the quantum momentum operator becomes $ \hat{p}_a= -i \frac{d}{da}$ while the quantum Hamiltonian (\ref{quham}) , Weyl ordered, reads:
\begin{equation}
\hat{H}= -\frac{1}{\sqrt{g[a,0]}}\frac{d}{d a}\left(\sqrt{g^{-1}[a,0]}\frac{d}{d a} \right) -V_{\rm eff}^{KA}[a,0] \label{quham}
\end{equation}
and the unbounded motion orbits become scattering backward waves while bound states arise from the bounded orbits only complying with tsome Bohr-Sommerfeld quantization conditions.

If both Kink and AntiKink are excited and we let the amplitude  $A$ vary things are different. The effective dynamics is captured by a Lagrangian system with two degrees of freedom: $a$ and $A$. Denoting now $a=a^1$ and $A=a^2$, the efective Lagrangian reads
\[
L=\frac{1}{2}g_{a^ia^j}[a^1,a^2]\frac{\partial a^i}{\partial \tau}\cdot \frac{\partial a^j}{\partial \tau } -V_{\rm eff}^{KA}[a^1,a^2] \qquad , \qquad i,j=1,2 \, \, .
\]
and, accordingly, the motion equations become:
\[
\frac{\partial^2 a^i}{\partial \tau^2}+ \sum_{j,k}\, \Gamma^{a^i}_{a^j a^k}\,\frac{\partial a^j}{\partial \tau}\frac{\partial a^k}{\partial \tau}=-\sum_j g^{a^ia^j} \frac{\delta V}{\delta a^j}  \quad , \quad \Gamma^{a^i}_{a^j a^k}=\frac{1}{2}\sum_l g^{a^i a^l}\left(\frac{\partial g_{a^l a^j}}{\partial a^k}+\frac{\partial g_{a^l a^k}}{\partial a^l}-\frac{\partial g_{a^k a^j}}{\partial a^l}\right)
\]
The energy is still a constant of motion but there is no a second invariant that would guarantee the Liouville integrability of the system.
Needless to say, to obtain analytical solutions of this system of ODE's is hopeless. Nevertheless, numerical analysis of this system for initial conditions corresponding to Kink/AntiKink scattering has been sucessefully performed in the seminal Reference \cite{MaOlRoWe}. These authors reached similar consequence to those previously obtained in the numerical treatment of the full field theory: in the Kink Antikink scattering quasi-bound states where several bounces occur arise for some windows of initial velocities. Moreover, the pattern shows  a very interesting fractal structure.

\subsection{Fermionic fluctuations of Kink/Anti-Kink configurations}

Consider the bounded spinorial fluctuation over the Kink/AntiKink configuration (\ref{fkasm})  when the ratio $\frac{g}{\lambda}$ is $2$.
\begin{eqnarray}
&& \frac{1}{\sqrt{g}}\Psi_{\sqrt{3}\sqrt{3}}(y,a,\Lambda)=\frac{\Lambda}{\tanh[a]}\times \label{fkasm} \\ && \times  \left[\sech(y+a)\left(\begin{array}{c} {}_2F_1(4,-1,2, \frac{1}{2}(1+\tanh(y+a))\\ {}_2F_1(3,0,2, \frac{1}{2}(1+\tanh(y+a)) \end{array}\right)- \sech(y-a)\left(\begin{array}{c}{}_2F_1(4,-1,2, \frac{1}{2}(1+\tanh(y-a))\\ {}_2F_1(3,0,2, \frac{1}{2}(1+\tanh(y-a)) \end{array}\right) \right] \nonumber
\end{eqnarray}

\begin{figure}[ht]
\centerline{\includegraphics[height=4cm]{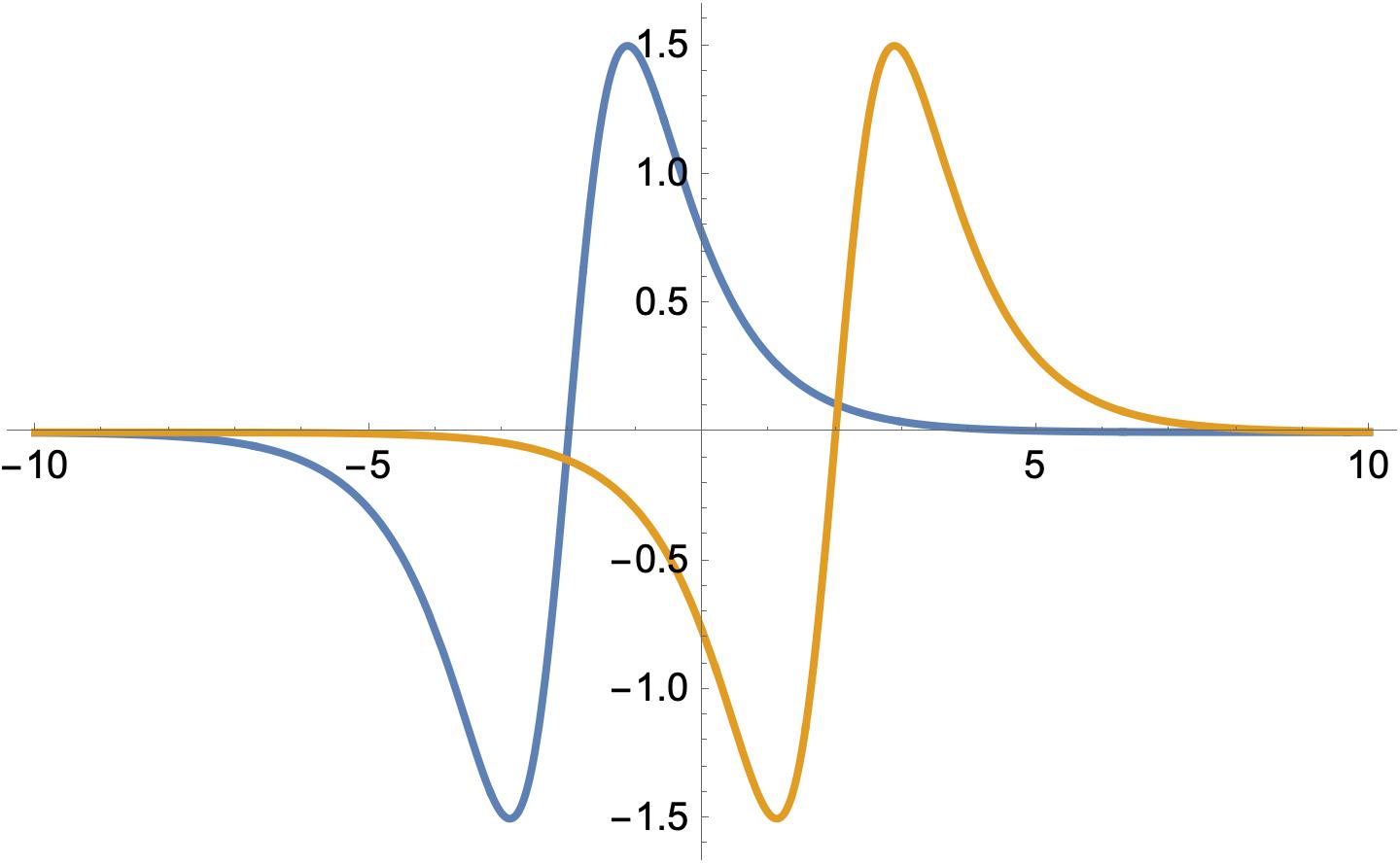} \hspace{2cm} \includegraphics[height=4cm]{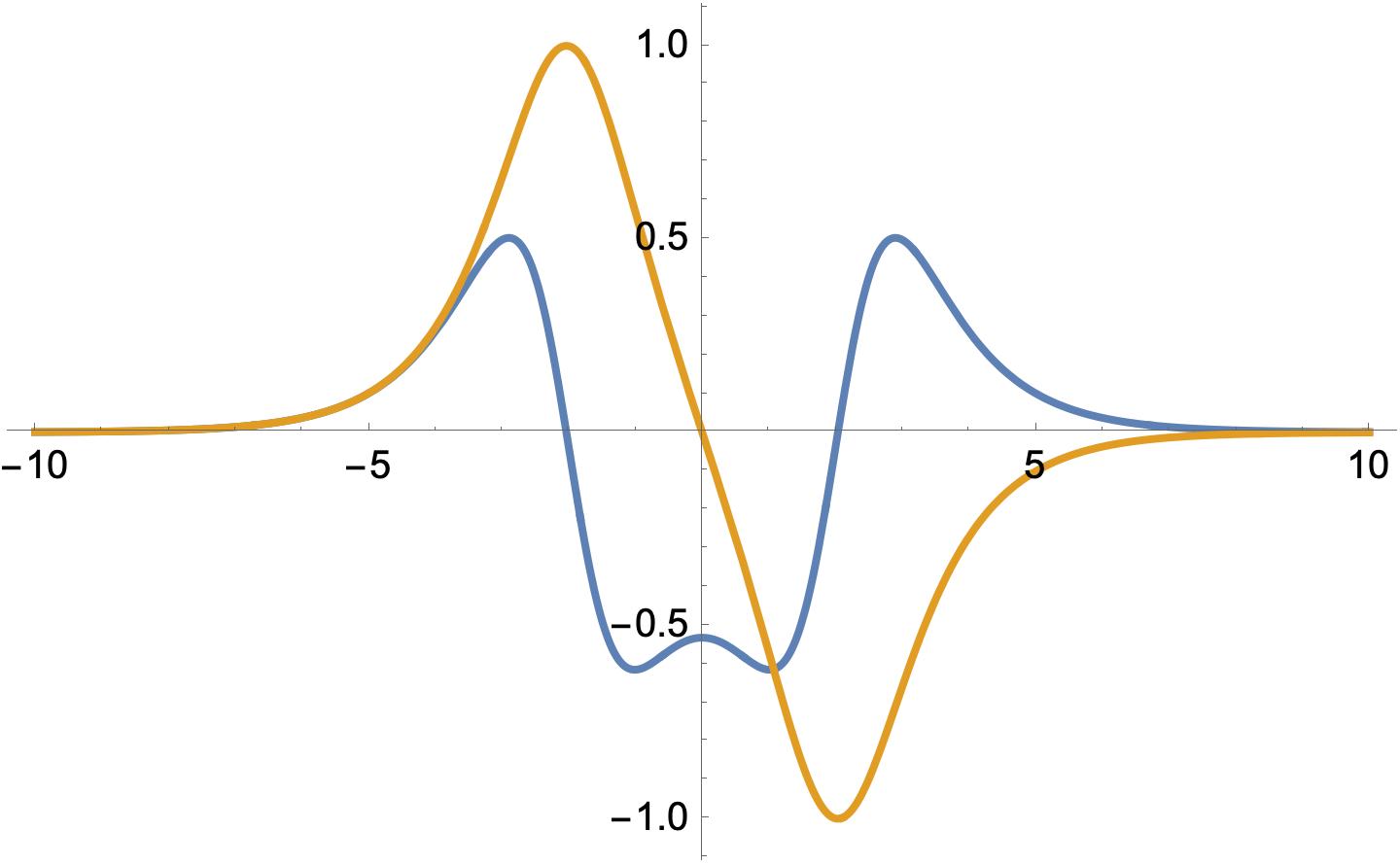}}
\caption{(left) Graphics of the Kink shape mode and the AntiKink shape mode (right) Graphics of the upper and lower component of the Fermionic shape mode}
\end{figure}

Contribution to the Kinetic energy of the Fermionic fluctuations is encoded in the factor $G(a)$ (\ref{fkinenKAK})
\begin{eqnarray}
T_{\rm eff}^F&=& - \int_{-\infty}^\infty \, dy \Psi^\dagger_{\sqrt{3}\sqrt{3}}(y,a,\Lambda)\dot{\Psi}_{\sqrt{3}\sqrt{3}}(y,a,\Lambda)= -\Lambda^*\dot{\Lambda} G(a) \nonumber \\ G(a)&=&\frac{1}{\tanh^2 a}\int_{-\infty}^\infty \, dy \, \left[\left(\frac{\tanh (y+a)}{\cosh(y+a)}-\frac{\tanh (y-a)}{\cosh (y-a)}\right)^2+\left(\sech(y+a)-\sech(y-a)\right)^2\right] \nonumber\\ &=&\frac{4}{3} (12 a-3 \sinh (2
   a)-3 \sinh (4 a)+\sinh (6
   a)) \coth ^2(a)
   \text{csch}^3(2 a) \label{fkinenKAK}
\end{eqnarray}
It is clear in Figure 5 (left) that the modification of the metric due to fermionic shape modes is stronger when Kink and anti-Kink are close to each other

\begin{figure}[ht]
\centerline{\includegraphics[height=4cm]{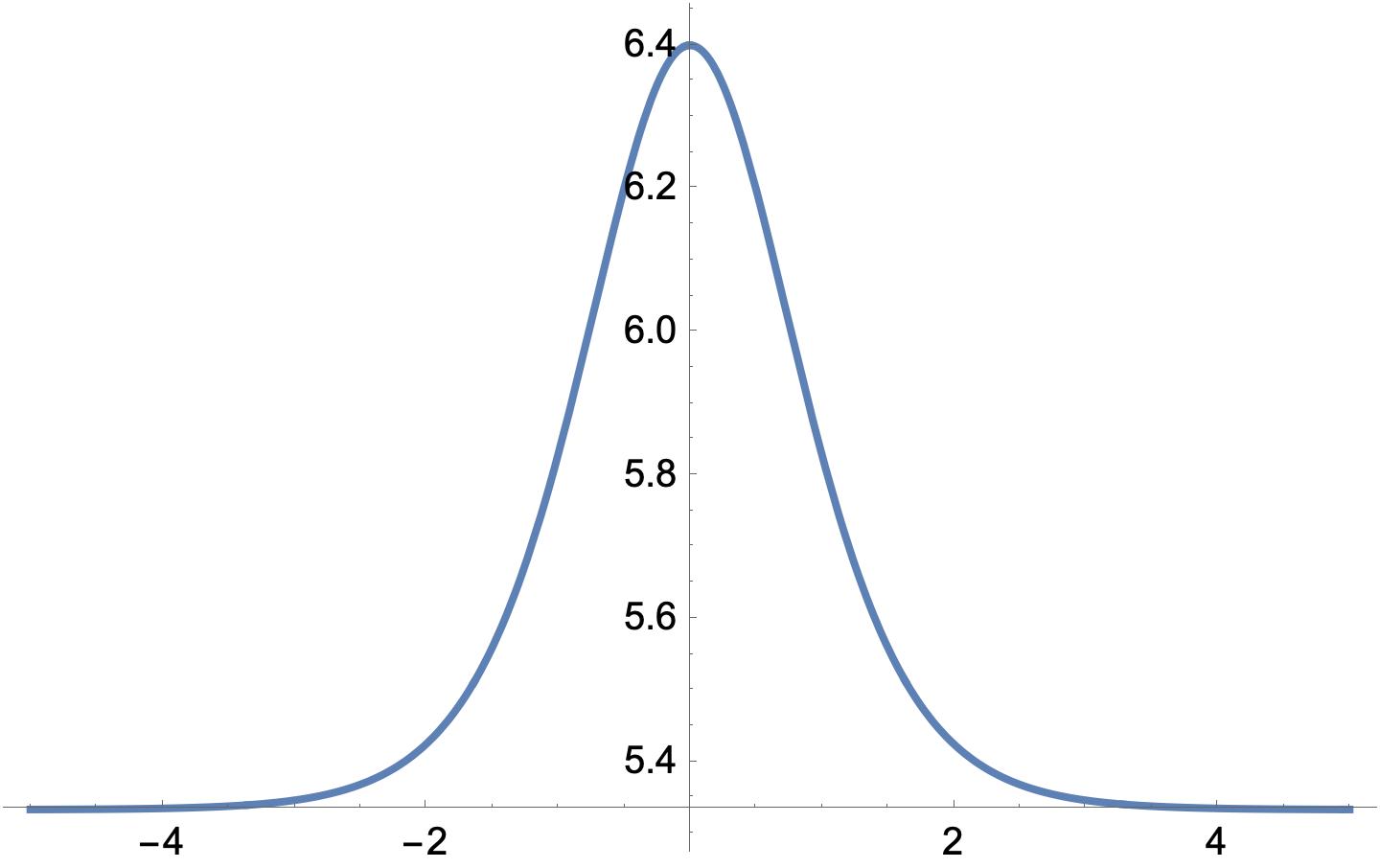} \hspace{2cm} \includegraphics[height=4cm]{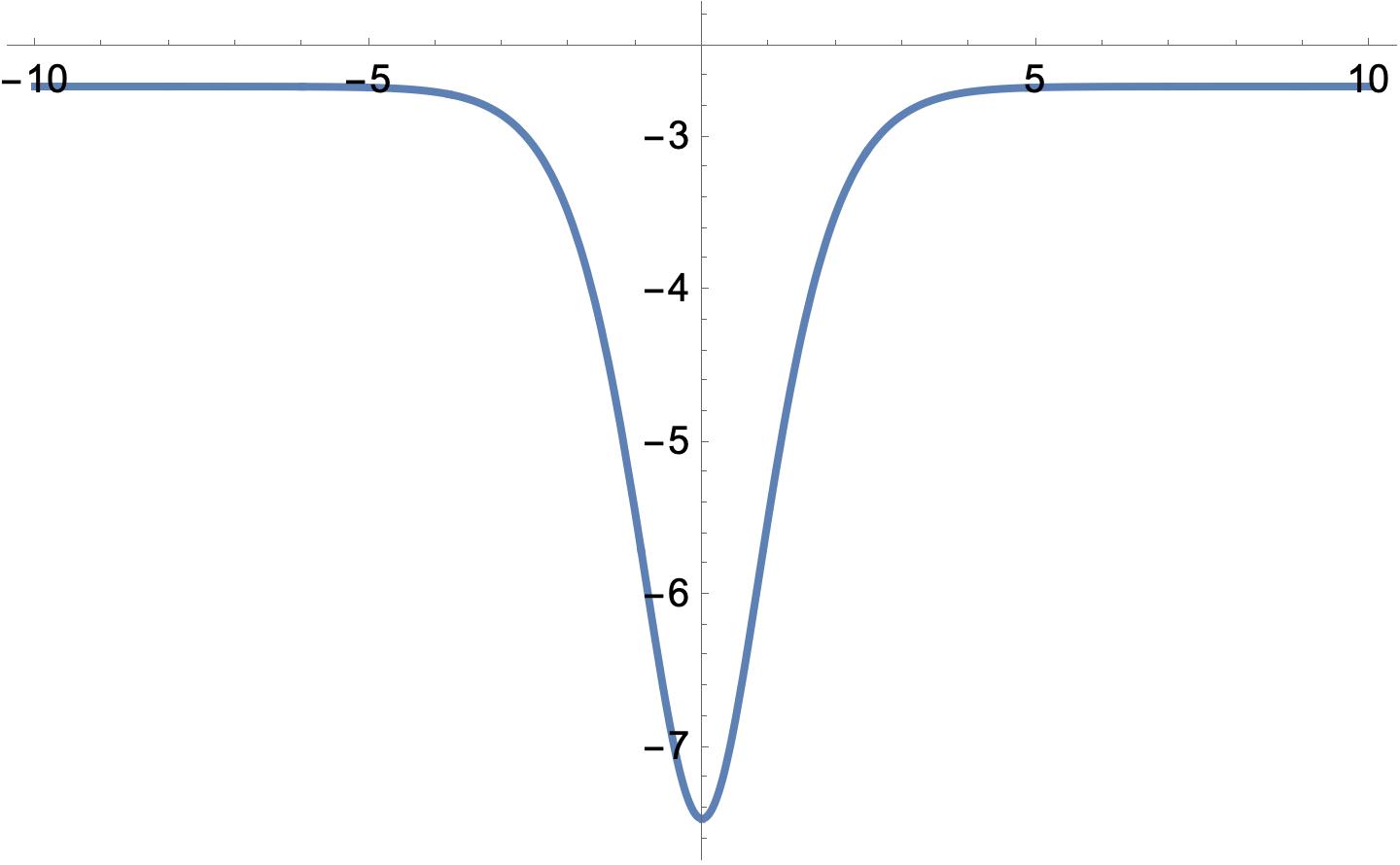}}
\caption{(left) Contribution to the matric of Fermionic fluctuations $G(a)$ as a function of the distance between Kink and Antikink (right) Effective potential  $U(a)$ due to Fermionic fluctuations of the Kink/AntiKink configuration also as a function of $a$.}
\end{figure}
Next we compute the efective potential contributed by the Fermionic shape mode fluctuating over the Kink/AntiKink configuration

\begin{eqnarray*}
  V_{\rm eff}^F(a,\Lambda^*\Lambda)&=&i \int_{-\infty}^\infty \, dy \, \Psi_{\sqrt{3}\sqrt{3}}^\dagger(y,a,\Lambda)\left\{-i \sigma_2 \frac{d}{d y}+2 \phi_{KA}(y,a,A) \right\} \Psi_{\sqrt{3}\sqrt{3}}(y,a,\Lambda)=i\Lambda^*\Lambda \cdot U(a)   \\ U(a)&=& 
  \frac{1}{\tanh^2 a}\int_{-\infty}^\infty \, dy \, \left[ \left(\frac{{}_2 F_1[4,-1,2,\frac{1}{2}(1+\tanh(y+a))]}{\cosh (y+a)}- \frac{{}_2 F_1[4,-1,2,\frac{1}{2}(1+\tanh(y-a))]}{\cosh (y-a)}\right)\times \right. \\ && \times  \left(-\frac{ d}{d y}+2 \phi_{KA}(y,a,A)\right)\left(\frac{{}_2 F_1[3,0,2,\frac{1}{2}(1+\tanh(y+a))]}{\cosh (y+a)}- \frac{{}_2 F_1[3,0,2,\frac{1}{2}(1+\tanh(y-a))]}{\cosh (y-a)}\right) + \nonumber \\&& + \left(\frac{{}_2 F_1[3,0,2,\frac{1}{2}(1+\tanh(y+a))]}{\cosh (y+a)}- \frac{{}_2 F_1[3,0,2,\frac{1}{2}(1+\tanh(y-a))]}{\cosh (y-a)}\right)\times \\ &&  \left. \times  \left(\frac{ d}{d y}+2 \phi_{KA}(y,a,A)\right)\left(\frac{{}_2 F_1[4,-1,2,\frac{1}{2}(1+\tanh(y+a))]}{\cosh (y+a)}- \frac{{}_2 F_1[4,-1,2,\frac{1}{2}(1+\tanh(y-a))]}{\cosh (y-a)}\right) \right]
\end{eqnarray*}
where $\phi_{KA}(y,a,A)$ is the excited Kink/AntiKink configuration defined in formula (\ref{KAvibad}). The result for the effective potential is (\ref{Feffpotka}):
\begin{equation}
U(a)= -\frac{2}{3} \coth ^2(a)
   \text{csch}^3(2 a)
   \left(\log \left(e^{36
   a}\right)-3 \sinh (2 a)-12
   \sinh (4 a)+\sinh (6 a)+6
   \log \left(e^{2 a}\right)
   \cosh (4 a)\right) \, \, \label{Feffpotka}
\end{equation}
In Figure 5(right) is clear that the effective potential also receives one (attractive) important contribution from the fermionic shape mode when tke Kink is close to the Anti-Kink.

Since the Fermionic Kink/ AntiKink fluctuations do not depend on the shape mode vibration amplitude of Kink and Antikink we expect that $G(a)$ only will be a function of $a$ and indeed this the case. The effective potential $U(a)$ induced by the Fermionic fluctuations on vibrating Kink AntiKink configurations however do not depend on the amplitude of the Bosonic shape mode . This unexpected effect is due to the fact, see Figure 8 (left), that Kink and Antikink oscillates in counter-phase and there are destructive interferences. Thus, one should expect that interactions between the amplitudes $\Lambda$ and $A$ respectively of Fermionic and Bosonic fluctuations of Kink/AntiKink configurations only would arise if the amplitudes of vibrations of Kink differ from the AntiKink amplitudes. To test this statement generalization to consider the amplitude of the Kink different from the AntKink amplitude in the shape modes is implemented
by replacing in the Kink AntiKink configuration the contribution of excitations by
\[
A\frac{\tanh(y+a)}{\cosh(y+a)}-B \frac{\tanh (y-a)}{\cosh (y-a)} \, .
\]
A quick Mathematica run confirms the conceptual argument given above:
\begin{eqnarray*}
U(a,A-B)&=&-\frac{\coth ^2(a)
   \left(\frac{3}{2} \pi 
   \left(e^{2 a}-1\right)^6
   (A-B)+16 e^{6 a} (36 a-3
   \sinh (2 a)-12 \sinh (4
   a)+\sinh (6 a)+12 a \cosh
   (4 a))\right)}{3 \left(e^{4
   a}-1\right)^3}
\end{eqnarray*}

Understanding of how the induced metric and effective potential by Fermionic Kink fluctuations depends on $\frac{g}{\lambda}$ starts 
by looking at $N=3$. In this case there are two Fermionic shape modes. The lower frequency is $\sqrt{5}$ that, implemented on
the excited Kink/AntiKink configuration, gives rise to the spinor wave function (\ref{lfspwa}):
\begin{equation}
\frac{1}{\sqrt{g}}\Psi_{\sqrt{5}\sqrt{5}}(y,\Lambda )= \frac{\Lambda}{\tanh a}\left(\begin{array}{c} 
\frac{{}_2 F_1[6,-1,3,\frac{1}{2}(1+\tanh(y+a))}{\cosh^2(y+a)}-\frac{{}_2 F_1[6,-1,3,\frac{1}{2}(1+\tanh(y-a))}{\cosh^2(y-a)}\\
\frac{{}_2 F_1[5,0,3,\frac{1}{2}(1+\tanh(y+a))}{\cosh^2(y+a)}-\frac{{}_2 F_1[5,0,3,\frac{1}{2}(1+\tanh(y-a))}{\cosh^2(y-a)}
\end{array}\right) \label{lfspwa}
\end{equation}
The effective kinetic energy induced by this Fermionic shape mode is
\begin{eqnarray}
T_{\rm eff}^{F1}&=&\int_{-\infty}^\infty \, dy \, \Psi_{\sqrt{5}\sqrt{5}}^\dagger(y,\Lambda)\dot{\Psi}_{\sqrt{5}\sqrt{5}}(y,\Lambda)=-\Lambda^*\dot{\Lambda}\cdot G3(a) \nonumber \\ G3(a) &=& \frac{1}{5} \coth ^2(a)
   \text{csch}^5(2 a)
   \left(-80 \sinh (2 a)-15
   \sinh (6 a)+\sinh (10
   a)+120 \log \left(e^{2
   a}\right) \cosh (2
   a)\right)  \label{metric31g}
\end{eqnarray}
where the $G3(a)$  factor is written in (\ref{metric31g}).

Likewise the effective potential is derived via similar procedures
\begin{eqnarray}
V_{\rm eff}^{F1}&=&i \int_{-\infty}^\infty \, dy \, \Psi_{\sqrt{5}\sqrt{5}}^\dagger(y,\Lambda) \{-i\sigma_2\frac{d}{dy}+3 \phi_{KA}(y,a,A)\}\Psi_{\sqrt{5}\sqrt{5}}(y,\Lambda)=i\Lambda^*\Lambda \cdot U3(a) \nonumber \\ U3(a)&=& -\frac{2}{15} \coth ^2(a)
   \text{csch}^5(2 a)
   \left(-350 \sinh (2 a)-125
   \sinh (6 a)+\sinh (10 a)+60
   \log \left(e^{2 a}\right)
   (11 \cosh (2 a)+\cosh (6
   a))\right) \nonumber
\end{eqnarray}

\begin{figure}[ht]
\centerline{\includegraphics[height=4cm]{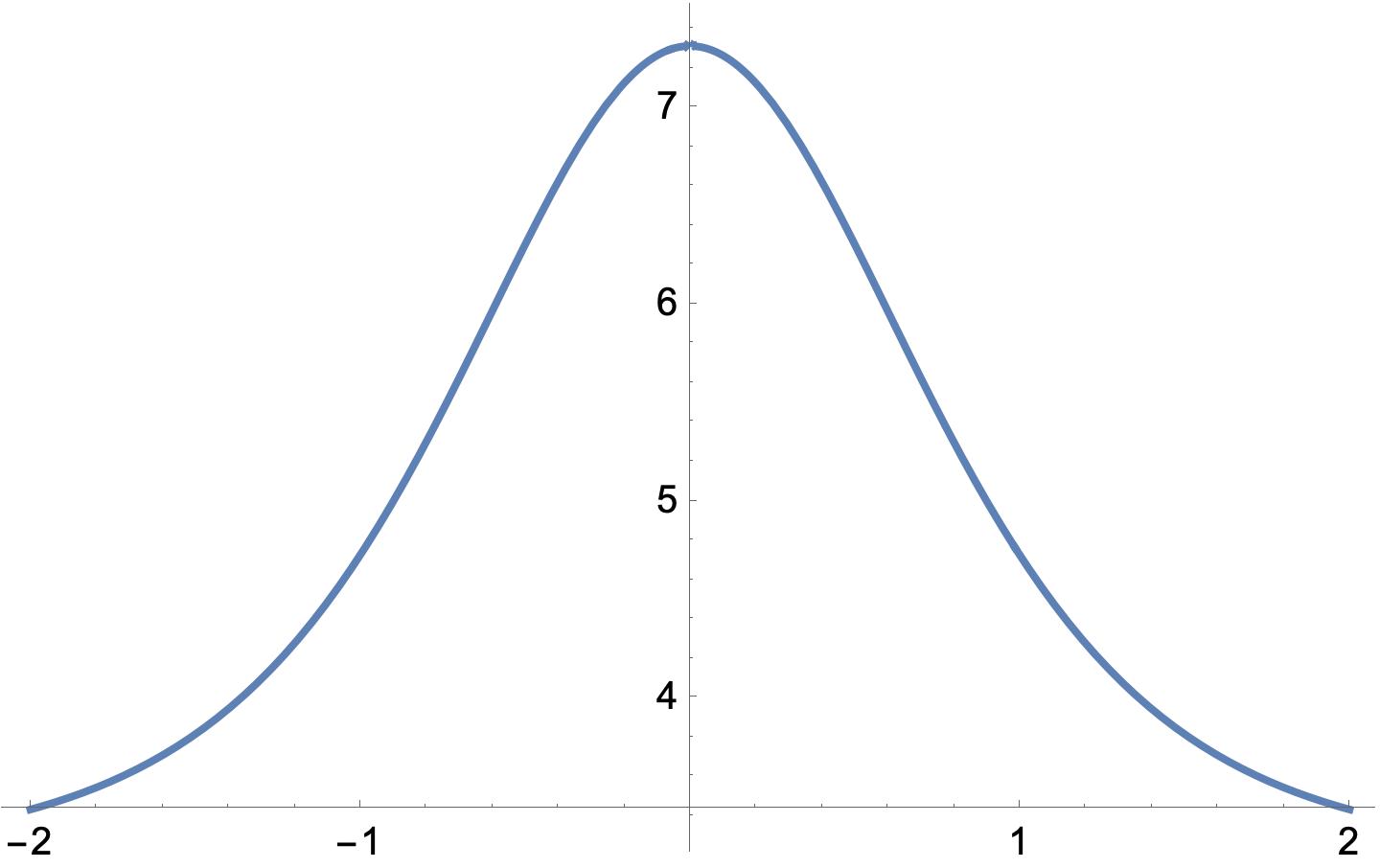}\hspace{2cm} \includegraphics[height=4cm]{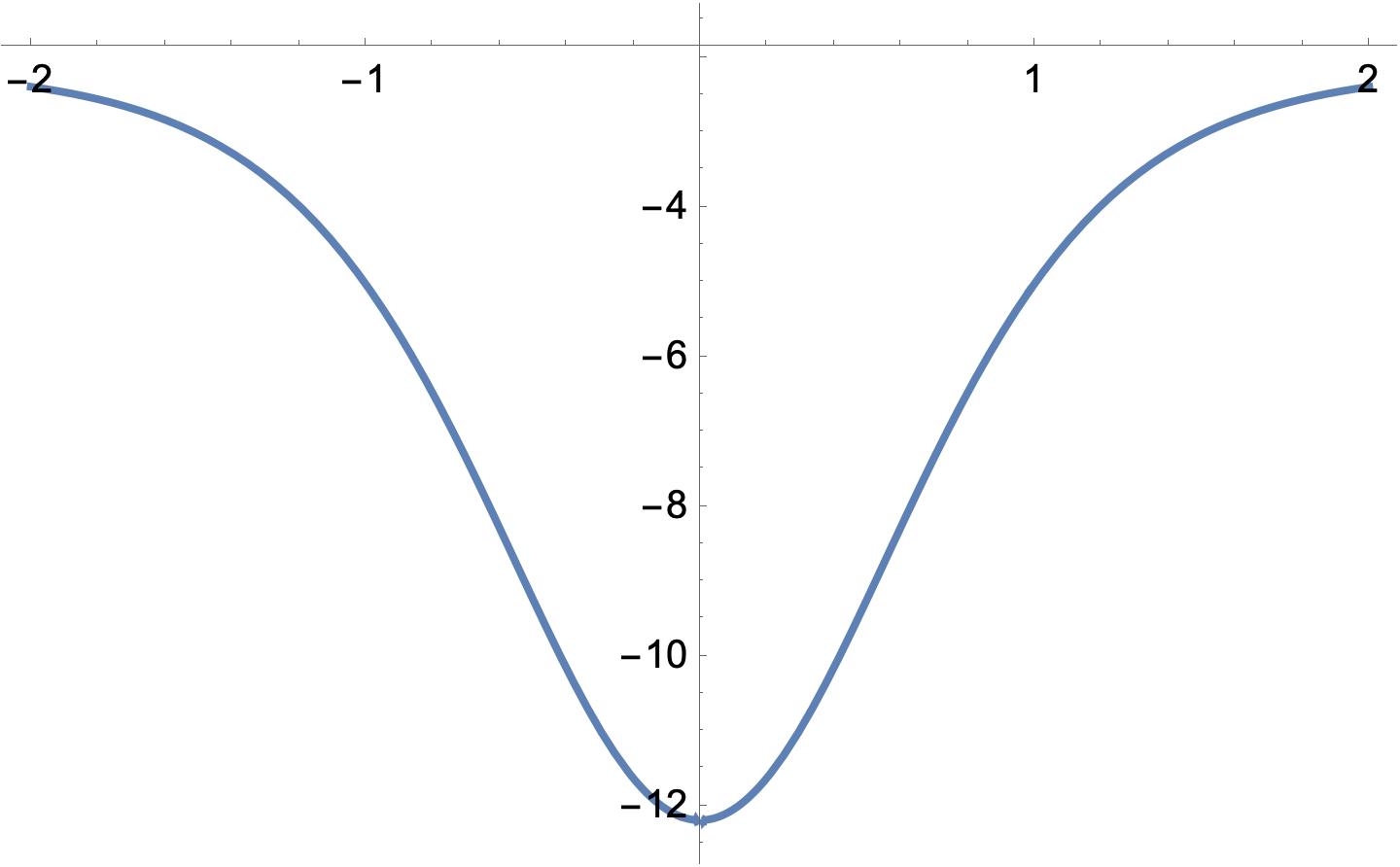}}
\caption{(left) Induced metric $G3(a)$ (right) Induced effective potential $U3(a)$  }
\end{figure}

In sum, identical pattern is observed in the effective adiabatic dynamics induced by Fermionic Kink shape modes in Kink/AntiKink
configurations when $N=2$ or $N=3$. We thus skip of writing the results obtained for the Fermionic shape mode of frequency $\sqrt{8}$
which are qualitatively equivalent but even more cumbersome.

\section{Outlook}
The theoretical developments in this work have been focused in a Field Theory $(1+1)$-dimensional model of bosons and fermions, specifically the Jackiw-Rebbi model introduced in Reference \cite{Jackiw}. One interesting way to extend these ideas is to jump in the number of fields both bosonic and fermionic. An early proposal in this line can be found in a paper published circa the year 2000
by one MIT group, see \cite{Jaffe}. In that work the authors deal with a Field Theory model with two Bose and two Fermi fields. The interaction between bosonic and fermionic fields is through two Yukawa couplings and sophisticated phenomena arise. We intend, however, 
to extend scalar/Bose two-field models treated by our group in Salamanca by incorporating  two spinor/Fermi fields in these systems. The first model that we have in mind was discussed, among other papers, in Reference \cite{Guilarte}. This Field Theoretical model exhibits a rich variety of Kinks and posseses interesting properties of integrability in the related mechanical analogous system. We expect that adding spinor/Fermi fields similar phenomena to that found in the Jackiw-Rebbi model will be kept but subtle novelties probably will be unfolded.
The second system where we envisage that the analysis developed in the JR model will be fruitful is the massive non-linear $\mathbb{S}^2$-sigma $1+1$-dimensional model, \cite{Guilarte1}. Again, in this \lq\lq deformed\rq\rq non-linear sigma model a manifold of Kinks, topological and non topological, exist. Spinor/Fermi fields will be included as sections of one spinor bundle over the two-sphere rather than spinor functions. In any case we expect that new phenomena will appear with respect to that described in the JR model.
The interplay between scalar and spinor fields in this non-linear system promises the appearance of new subleties.

Other playground where the analysis of effective low energy dynamics seems to be promising is the moduli space of BPS vortices in the Abelian Higgs model, see e.g. \cite{Queiroga} for a parallel study devoted to Kinks. The Abelian Higgs model in $(2+1)$ space-time, at the critical ratio between the 
self-interacting scalar $\lambda$ and the electromagnetic $e^2$ couplings, the transition point between Type I and Type II Ginzburg-Landau superconductivity, posseses manifolds of topological non interacting stable BPS vortices characterized by an integer number of magnetic quanta. 
These topological defects admit also vibrational modes, see References  \cite{Alonso4}-\cite{Alonso5}. The collective coordinate low energy analysis has been developed in \cite{Manton}-\cite{Manton1}-\cite{Manton2} in absence of Fermions. The addition of Fermions 
to the Abelian Higgs model is compelling, including Yukawa and electromagnetic couplings. The system will be much more complex but the temptation is strong towards identify the Collective Coordinates of these planar topological defects and see how Fermions affect the adiabatic dynamics of BPS vortices.

\end{document}